\documentclass[english,12pt]{article}
\usepackage[utf8]{inputenc}

\usepackage[left=2.0cm, right=2.0cm, top=2.5cm, bottom=2.5cm]{geometry}

\usepackage{mathrsfs}
\usepackage{ dsfont }

\makeatletter

\usepackage{hyperref}
\hypersetup{
	colorlinks,
	citecolor=blue,
	filecolor=blue,
	linkcolor=blue,
	urlcolor=blue,
	linktocpage=true
}

\usepackage{authblk}
\usepackage{lmodern}
\usepackage{amsmath}
\usepackage{amsfonts}
\usepackage{amssymb}
\usepackage{amsthm}
\usepackage{graphicx}
\usepackage{url}
\usepackage{subcaption}
\usepackage{array}
\usepackage{cancel}
 \usepackage{pdfpages}
\usepackage{autobreak}
\usepackage{floatrow}
\usepackage{booktabs}
\usepackage{bbm}
\usepackage{cite}

\usepackage{tikz}
\usetikzlibrary{shapes.geometric, arrows}
\tikzstyle{io} = [trapezium, trapezium left angle=70, trapezium right angle=110, minimum width=3cm, minimum height=1cm, text centered, draw=black, fill=blue!30]
\tikzstyle{startstop} = [rectangle, rounded corners, minimum width=3cm, minimum height=1cm,text centered, draw=black]
\tikzstyle{contra} = [rectangle, minimum width=3cm, minimum height=1cm, text centered, draw=red, fill=orange!30]
\tikzstyle{decision} = [diamond, minimum width=3cm, minimum height=1cm, text centered, draw=black, fill=green!30]
\tikzstyle{arrow} = [thick,->,>=stealth]

\renewcommand{\arraystretch}{1.4}

\usepackage{booktabs}
\newcommand{\ra}[1]{\renewcommand{\arraystretch}{#1}}

\newcommand{\iu}{{i\mkern1mu}}
\newcommand{\C}{{\mathcal C}}
\newcommand{\g}{{\gamma}}
\newcommand{\td }{ \text{d} }
\newcommand{\Z }{ \tilde{Z} }
\newcommand{\LL}{{\sf \Lambda}}
\newcommand{\nwDelta}{{\varrho }}
\newcommand{\nwl}{{t}}
\newcommand{\uu }{{ \mathfrak{u}(1)}}

\newcommand{\lb}{\left( }
\newcommand{\rb}{ \right) }

\newcommand{\llangle }{\langle \langle }
\newcommand{\rrangle }{ \rangle \rangle }

\newcommand{\R}{\ensuremath{\mathbb R}}

\theoremstyle{plain}
\newtheorem{lemma}{Lemma}[section]

\theoremstyle{definition}

\newcommand{\normord}[1]{:\mathrel{#1}:}

\title{Fake Z}

\author[1]{Anatoly Dymarsky}
\author[2]{Rohit R. Kalloor}
\affil[1]{Department of Physics and Astronomy\\University of Kentucky\\Lexington, KY, USA 40506}
\affil[2]{Department of Particle Physics and Astrophysics\\Weizmann Institute of Science\\Rehovot 7610001, Israel}

\begin{document}

\newcommand\barparen[1]{\overset{(-)}{#1}}

\maketitle

\begin{abstract}
    
Recently introduced connections between quantum codes and Narain CFTs provide a simple ansatz to express a modular-invariant  function $Z(\tau,\bar \tau)$ in terms of a multivariate polynomial satisfying certain additional properties. 
These properties include  algebraic identities, which ensure modular invariance of $Z(\tau,\bar \tau)$, and positivity and integrality of coefficients, which  imply positivity and integrality of the $\mathfrak{u}(1)^n \times \mathfrak{u}(1)^n$ character expansion  of $Z(\tau,\bar \tau)$. Such polynomials 
come naturally from codes, in the sense that each code of a certain type gives rise to the so-called enumerator polynomial, which automatically satisfies all necessary properties, while the resulting $Z(\tau,\bar \tau)$ is the partition function of the code CFT  -- the Narain theory unambiguously constructed from the code. Yet there are also ``fake'' polynomials satisfying all necessary properties, that are not associated with any code. They lead to  $Z(\tau,\bar \tau)$ satisfying all modular bootstrap constraints (modular invariance and positivity and integrality of character expansion), but whether they are partition functions of any actual CFT is unclear. We consider the group of the six simplest fake polynomials and denounce the corresponding $Z$'s as fake: we show that none of them is the torus partition function of any Narain theory. Moreover, four of them are not partition functions of any unitary 2d CFT; our analysis for other two is inconclusive. Our findings point to an obvious limitation of the modular bootstrap approach: not every solution of the full set of torus modular bootstrap constraints is due to an actual CFT.  In the paper we consider six simple examples, keeping in mind that thousands more can be constructed. 
\end{abstract}

\tableofcontents

\newpage

\section{Introduction}
Conformal modular bootstrap \cite{Hellerman,hellerman2011bounds,keller2013modular,friedan2013constraints,qualls2014bounds,
hartman2014universal,qualls2015universal,kim2016reflections,lin20172,anous2018parity,collier2018modular,afkhami2019fast,
cho2019genus,hartman2019sphere,afkhami2020high,afkhami2021free,gliozzi2020modular}, leverages consistency conditions of 2d CFT torus partition functions $Z(\tau,\bar \tau)$, and has  proved to be a powerful tool yielding many new interesting results about 2d theories. Similar to its higher-dimensional cousin: the conformal bootstrap \cite{rattazzi2008bounding,poland2016conformal,poland2019conformal}, the modular bootstrap is merely a set of necessary conditions, which any 2d CFT must satisfy. Thus solving modular bootstrap constraints, in the sense of finding appropriate $Z(\tau,\bar \tau)$, does not guarantee one is dealing with an actual CFT. The same of course applies to any version of bootstrap. So far this has not been a problem partially because the focus was on the exclusion plots, charting the space of CFT data not compatible with bootstrap constraints. Besides, we have been lucky (though the reason for the ``luck'' is unclear) in the sense that points sitting at the edge of exclusion plots (solutions of certain bootstrap constraints) happen to be actual CFTs, as is the case of 3d Ising model \cite{el2012solving}. Yet it would be naive to take for granted that a solution to modular bootstrap constraints is always an actual \textit{bona fide} CFT. To drive the point home, in this paper we consider several explicit examples of would-be torus partition functions $Z(\tau,\bar \tau)$ satisfying all apparent properties, such as modular invariance and decomposition into characters with positive-integer coefficients, yet we show these are not partition functions of any 2d CFT.

These examples come from the relation between Narain CFTs and quantum codes and were first formulated in \cite{ds,dymarsky2021solutions}. Each code $\C$ (quantum stabilizer code of length $n$, and of the self-dual, real type) gives rise to a refined enumerator polynomial $W_\C(x,y,z)$, a homogeneous polynomial of degree $n$ satisfying several additional properties: i) all coefficients are positive integers ii) $W_\C(1,0,0)=1$, and iii)
\begin{eqnarray}
\label{Real}
W_\C(x,y,z)&=&W_\C(x,-y,z),\\
W_\C(x,y,z)&=&\tfrac{1}{2^n} W_\C(x+y+2z,x+y-2z,x-y).
\label{McW}
\end{eqnarray}
Starting from $\C$ one can construct a Narain CFT with central charge $n$ and torus partition function given by  (extension to higher-genus partition functions was recently discussed in \cite{henriksson2021codes,henriksson2022narain})
\begin{align}
Z_\C \lb \tau , \bar{ \tau } \rb 
	& = 
		\frac{ W_\C \lb \theta _3 \bar{\theta }_3 + \theta _4 \bar{\theta} _4 , \theta _3 \bar{\theta }_3 - \theta _4 \bar{\theta} _4  , \theta _2 \bar{\theta }_2 \rb }{ \lb 2 \eta \lb \tau \rb  \bar{ \eta } \lb \bar{ \tau } \rb \rb ^n },
		\label{eqn:recipe}
\end{align}
where $\theta_2,\theta_3,\theta_4(\tau)$ are the standard Jacobi theta functions. 
It is straightforward to see that \eqref{Real} and \eqref{McW} ensure invariance of $Z_\C $ under $\tau\rightarrow \tau+1$ and $\tau\rightarrow -1/\tau$ respectively; condition ii) ensures $Z_\C$ has unique vacuum state, and i) guarantees $Z_\C$ can be decomposed into $\mathfrak{u}(1)^n\times \mathfrak{u}(1)^n$ characters with positive-integer coefficients. In other words, connection with codes provides an ansatz to solve all torus modular bootstrap constraints in terms of a polynomial satisfying certain algebraic identities. This approach was subsequently generalized to include codes and Narain CFTs of the more general type \cite{buican2021quantum,Yahagi_2022,angelinos2022optimal}, while some of the main features remain the same: the partition function $Z_\C(\tau,\bar \tau)$ is written in terms of a multivariate polynomial satisfying certain algebraic identities. This form implies $Z$ is a sum of a finite, although potentially large, number of  ``characters'', though this description is schematic because the relation with codes includes non-rational CFTs \cite{dymarsky2021non,angelinos2022optimal}. The algebraic approach stemming from codes is complementary to the approach to construct modular invariant $Z$'s by combining only a few characters \cite{Gaberdiel_2016,Chandra_2019,Chandra2019,mukhi2019classification}. 

While each code gives rise to a $W_\C$, there could be polynomials satisfying all the necessary conditions, yet not associated with any code. This is a completely general situation for codes of any type, e.g.~classical, quantum, etc. For the small code length $n$ it is straightforward to classify which polynomials are ``genuine'' (code-related), and which ones are ``fake'' (not associated with any code) because all codes and all polynomials can be found explicitly. For large $n$ this problem is highly non-trivial.\footnote{This is related to the problem of understanding for which $n$ optimal codes are extremal.} For the polynomials of the type introduced above, all polynomials with $n=1,2$ are genuine, but starting from $n\geq 3$ there is a rapidly growing number of fake polynomials.  For $n=3$, there are six fake polynomials  \cite{dymarsky2021solutions},
\begin{align}
W_1 \lb x , y , z \rb & = 
x^3 + 2 x^2 z + y^2 z + 3 x z^2 + z^3 \nonumber \\
W_2 \lb x , y , z \rb & = 
x^3 + x^2 z + 2 y^2 z + 3 x z^2 + z^3 \nonumber \\
W_3 \lb x , y , z \rb & = 
x^3 + x y^2 + 2 x^2 z + 2 x z^2 + 2 z^3 \nonumber \\
W_4 \lb x , y , z \rb & = 
x^3 +  x y^2 + 2 y^2 z + 2 x z^2 + 2 z^3 \nonumber \\
W_5 \lb x , y , z \rb & = 
x^3 + 2 x y^2 + x^2 z + x z^2 + 3 z^3 \nonumber \\
W_6 \lb x , y , z \rb & = 
x^3 + 2 x y^2 + y^2 z + x z^2 + 3 z^3
\label{eqn:fake}
\end{align}
leading to six ``impostor'' partition functions $Z_i$ via \eqref{eqn:recipe}. The ``impostor'' status here indicates that they are not directly associated with any 2d CFT, but whether such a CFT could potentially exist is \textit{a priori} unclear.
We analyze the $Z_i$ to show that none of them can be the partition function of a Narain theory, and confirm that last four with $3\leq i \leq 6$ are fake -- they are not partition functions of {\it any} unitary 2d CFT. Our analysis, based on the peculiarities of chiral states, is inconclusive for $i=1,2$. 

Our analysis consists of two logically independent parts. The structure of an impostor partition function given by \eqref{eqn:recipe} is consistent with the $\mathfrak{u}(1)^n \times \mathfrak{u}(1)^n$ chiral algebra, hence one may suspect $Z$ is the partition function of a Narain theory. Our first step is to rule that out. We notice that \eqref{eqn:recipe} with the polynomials \eqref{eqn:fake} yields spectra inconsistent with the linear structure of the OPE coefficients. Namely, all six would-be partition functions $Z_i$, interpreted as the partition functions of Narain theories, include  $\mathfrak{u}(1)^n \times \mathfrak{u}(1)^n$ primary states with the dimensions $\Delta\leq 1$ and zero spin $\ell=0$. They also include states with $\Delta=7/4$ which all have spin $\ell=\pm 2$. In a  Narain CFT, each primary is a lattice vector and hence their linear combination with integer coefficients must be a primary as well. Yet, we show a particular combination of $\ell=0$ states must have zero spin and $\Delta=7/4$, leading to a contradiction. This argument can be put on a more general footing, suitable for generalizations.  
Any $Z$ obtained via \eqref{eqn:recipe} yields the dimensions of 
$\mathfrak{u}(1)^n \times \mathfrak{u}(1)^n$ primaries which are quantized in units of $1/4$, namely $\Delta=k/4$ for some integer $k$. In terms of corresponding Narain lattices, this means all lattice vectors are located half-integer distance-squared away from the origin, i.e., $r^2=k/2,\, k\in \mathbbm{Z}$. All Euclidean low-dimensional lattices of this kind can be constructed explicitly. Apparently there are no lattices with the theta series (spectrum) matching $Z_1,Z_2$ and unique lattices matching $Z_i$ for $i=3,4,5,6$. Here we rendered $Z_i$ ``Euclidean'' by considering pure imaginary $\tau$. To extend the comparison for arbitrary $\tau$ we need to equip Euclidean lattices with a Lorentzian metric. Apparently no metric exists to reproduce $Z_i$ for $i=3,4,5,6$ with arbitrary $\tau$. This simply means that no $Z_i$ is the partition function of a genuine Narain theory. 

The second argument is based on the analysis of chiral states. Assuming that a given $Z$ is the partition function of a genuine CFT, we analyze chiral states with $\Delta=\ell$ and list possible chiral algebras consistent with this spectrum. Then, for each such scenario we find that particular chiral algebras require additional states in the spectrum, beyond those present in $Z$. Using contradictions of this kind, we rule out all possible scenarios one by one for $Z_i$ for $i=3,4,5,6$. For $i=1,2$ our analysis is inconclusive, but in principle can be extended. 
The logic above rules out unitary 2d CFTs of any kind, thus rendering first argument unnecessary, at least for $i=3,4,5,6$. We nevertheless present both arguments because they are completely independent and rely on two  different techniques. Besides, our second approach falls short of yielding a conclusive result for $i=1,2$.

The paper is organized as follows.
Section \ref{NTintro} recalls some general facts about Narain CFTs and also sets the notation. In Section \ref{sec:elementary}, we give a proof in elementary terms that $Z_3$ is not a partition function of a Narain theory. Similar proof for other $Z_i$ are delegated to appendix \ref{elementaryproof}. A more systematic treatment of the Narain case is given in section \ref{alllattices}, where we construct the list of all Euclidean lattices which can potentially give rise to $Z_i$, and show $Z_{1,2}$  don't come from the Narain CFTs. To rule out  other $Z_i$, we consider equipping these Euclidean lattices with the Lorentzian structure, which is discussed in section \ref{sec:lor}. This completes the case of ruling out Narain theories. We give a general argument based on peculiarities of chiral algebra for $Z_{3,4,5,6}$ in section \ref{sec:final}. We conclude in section \ref{conclusions}.

\section{``It's not Narain''}

In this section, we show that none of the impostor functions $Z_i(\tau,\bar \tau)$, defined by \eqref{eqn:recipe} with help of  \eqref{eqn:fake},
 originate from the CFTs of the Narain type. 

\subsection{Notations and preliminaries}
\label{NTintro}
A Narain CFT is a theory of non-interacting bosons that move on a $n$-dimensional torus, and are coupled to a background $B$-field:
\begin{align}
S & = \frac{1}{4 \pi \alpha ' } \int \td ^2 x \lb  \partial _{\xi} X_i \partial ^{\xi} X^i + \epsilon^{\xi \zeta} B_{ij} \partial _{\xi} X^i  \partial _{\zeta}  X^j \rb
\end{align}
where in what follows we set $\alpha ' = 2$. Here $B_{ij}$ is an antisymmetric $n \times n $ matrix and the $ X^i $ (with $ i =1, ... , n $) are periodic: $ X^i \sim X^i + f^i $, for any element $\overrightarrow{f}$ of an $n-$dimensional lattice $\Gamma $. The lattice $\Gamma $ and the $B$-field characterise the space of Narain CFTs.\footnote{This parametrisation is rife with degeneracies, with multiple $ (\Gamma , B)$'s mapping to the same theory.}

Any such theory has central charge $c = \bar{c} = n$, and admits a $\mathfrak{u}(1)^n \times \mathfrak{u}(1)^{n}$ current algebra (throughout the rest of this section, we'll just be calling it the $ \mathfrak{u}(1)$ current algebra) generated by:
\begin{align}
\quad  j ^a (z) = \iu \partial  X ^a ,\quad  \bar{j} ^a (\bar{z}) = \iu \bar \partial  X ^a, \qquad a = 1, ... , n 
\end{align}
 In special cases (which will be relevant to us later), this current algebra is enhanced. 
 
The operator spectrum of the theory may then be organised into $ \mathfrak{u}(1)$ primaries and their descendants. The former are vertex operators of the form:
\begin{align}
\mathcal{V}_{\alpha, \beta} = e^{ \frac{\iu}{\sqrt{2}} ( \alpha _j X^j + \beta _j \bar{X}^j ) }
\end{align}
where
\begin{align}
X  = X_L + X_R,\qquad \bar{X} = X_L - X_R, \nonumber
\end{align}
and the momenta $ (\alpha, \beta )$ form a lattice $\LL$ generated by:
\begin{align}
\label{LambdaBg}
\Lambda & = 	\frac{1}{\sqrt{2}}
				\begin{pmatrix}
					2 \gamma ^* & B \gamma \\
						0 		& \gamma 
				\end{pmatrix} 		
\end{align}
where $\gamma $ generates $\Gamma $. It is easy to verify that
\begin{eqnarray}
\Lambda ^t \eta \Lambda & = \eta ,\qquad \eta
			= \begin{pmatrix}
				0 & \mathbbm{1}_n \\
				\mathbbm{1}_n & 0
				\end{pmatrix},
\label{eqn:esd}
\label{eqn:Narain}
\label{eta}
\end{eqnarray}
and therefore the set of $\mathfrak{u}(1)$ primaries has the structure of a $(n,n)$-dimensional Lorentzian lattice that is even and self-dual with respect to the metric $\eta$.
In other words, $\LL$ is a \textit{Narain} lattice. The mapping between the set of primaries and $\LL$ is natural:
\begin{align}
\mathcal{V}_{\alpha_1, \beta_1 } \times \mathcal{V}_{\alpha_2, \beta_2 } \sim \mathcal{V}_{\alpha_1 + \alpha_2, \beta_1 + \beta _2 } 
\end{align}
 where the right-hand side is the leading contribution to the OPE.

The torus CFT partition function of the theory can also be organised into contributions from $\mathfrak{u}(1)$ towers:
\begin{align}
Z & = \frac{1}{|\eta|^{2n}}  \sum_{(\alpha , \beta ) \in \LL} q^{(\alpha + \beta )^2 / 4} {\bar q}^{(\alpha - \beta )^2/ 4}
\end{align}
where
\begin{align}
\quad q=e^{2\pi i \tau}= \varrho \ t ,\quad \bar q=e^{-2\pi i \bar \tau} = \varrho \ t^{-1}. 
\end{align}
The \textit{vertex partition function} $\tilde Z\equiv Z\, |\eta|^{2n}$ in the variables $(\varrho, t )$ gives the combined Siegel theta function of $\LL$,
\begin{align}
{\tilde Z} 
	& = 
1 + \sum_{(\Delta, \ell) \neq 0}\, C_{\Delta,\ell} \varrho ^{\Delta} t^{\ell}
	= 
	\sum _{v = (\alpha ,\beta ) \in \LL } \varrho ^{ \frac{1}{2} v^t v } t^{\frac{1}{2} v^t \eta v }
	\label{eqn:thetaExp}
\end{align}
where $C_{\Delta,\ell} \in \mathbbm{Z}_{\geq 0}$ counts the number of primaries with dimension $\Delta$ and (signed) spin $\ell$. Hence, for a lattice point $ v = (\alpha , \beta ) \in \LL$, the Euclidean and Lorentzian norms (respectively) give the dimension and spin of the corresponding vertex primary:
\begin{align}
\Delta _v & = \frac{1}{2} v^t v =  \frac{1}{2} ( \alpha^2 + \beta^2 ) 
\nonumber \\
\ell _v & = \frac{1}{2} v^t \eta v =  \alpha . \beta 
\label{eqn:dimSpin}
\end{align}

To summarize, the Narain theory is defined by a Narain lattice $\Lambda$ with the $ \mathfrak{u}(1)$ primaries being in one to one correspondence with the lattice vectors. 

In what follows, we  will discusses lattices without explicitly specifying generating matrices or even the form of Lorentzian norm $\eta$, which could be differ from \eqref{eta} by a $O(2n)$ orthogonal transformation.  
We introduce the notion of Euclidean and Lorentzian norms defined for any vector $v\in \LL \subset \R^{2n}$,
\begin{align}
v^t v & =v.v=   ||v ||^2,
\nonumber \\
v^t \eta v & =<v,v>=  \llangle v \rrangle ^2.
\end{align} 
Length in what follows would refer exclusively to Euclidean norm and sq-length to norm squared.

\subsection{Elementary proof for \texorpdfstring{$Z_3$}{Z3} }
\label{sec:elementary}

%
%

A simple observation, which generalizes to all $Z$'s obtained via \eqref{eqn:recipe}, is that rescaling $\rho\rightarrow \rho^8, t\rightarrow t^8$ renders $\tilde{Z}$  a  sum of even-integer powers of $\rho$ and $t$, 
\begin{align}
\tilde{Z}_3 ((\varrho t)^8 , (\varrho t ^{-1} )^8 )
        & = 
        \sum_{w \in 2 \LL _3 } \varrho ^{ ||w||^2 } t^{ \llangle w \rrangle ^2 }
        \nonumber \\
	& = 
		1 + 4 \varrho ^2  + 8 \varrho ^4 + 16 \varrho ^6 + \dots + 48 \lb  t^8+t^{-8} \rb \varrho ^{14} + \dots   
  \label{eqn:Z3}
\end{align}
This means that  if $\tilde{Z}_3$ is a Siegel theta function of some ``progenitor''  Narain lattice $\LL _3$, 
then the rescaled lattice $2 \LL_3$ would be an  even (and hence integer) lattice in the Euclidean sense, 
\begin{equation}
|| w ||^2 \in 2 \mathbbm{Z},\quad \forall w\in 2 \LL_3.
\end{equation}
The rescaling is not a crucial step and is done to simplify the following presentation. 
The mapping between lattice points of  $2 \LL_3$  and $ \uu $ primaries is now as follows, 
\begin{align}
    \Delta _w  = \frac{1}{8} || w ||^2,  \qquad 
    \ell _w  = \frac{1}{8} \llangle  w \rrangle ^2.
    \label{eqn:newDLreln}
\end{align}

Our goal will be  to force a contradiction between the given Euclidean and Lorentzian structures. We do this by showing that the sublattice generated by the scalar operators with $ \Delta < 1 $ must contain a scalar of Euclidean sq-length (norm  squared) 14, or $ \Delta = \frac{7}{4}$, which is absent in \eqref{eqn:Z3}.
We start by observing there are four vectors of sq-length 2. At least two of these (say $a_1, a_2$) must be linearly independent, the remaining two are then $ -a_1 , -a_2 $. Now, the following equation is called the polarisation identity:

\begin{align}
 || a_1 + a_2 ||^2 + || a_1 - a_2 ||^2 & = 2 ( ||a_1||^2 + ||a_2||^2 ) \nonumber \\ 
                               & = 8 \nonumber \\
                               & = 4 + 4 
                               \label{eqn:polaa}
\end{align}
which is to say
\begin{align}
 || a _1 + a _2 ||^2 = || a _1 - a _2 ||^2  = 4
\end{align}
This is the only split possible since all sq-lengths must be even integers, and only $ \pm \{ a_1, a_2 \} $ can have sq-length 2. This translates to $a_1 . a_2= 0$.

We now turn our attention to the sq-length 4 vectors;  the only (integral) linear combinations of $a_{1,2}$ with sq-length 4 are $ \pm a_1 \pm a_2 $. Hence the remaining sq-length 4 vectors are linearly independent -- let's pick one of these and call it $b_1$. Let's see how $b_1$ plays with $a_{1,2}$ by applying the polarisation identity:
\begin{align}
|| a_{1,2} + b_1 ||^2 + || a_{1,2} - b_1 ||^2 & = 2 ( ||a_{1,2}||^2 + ||b_1||^2 ) \nonumber \\
                               & = 12   \nonumber \\
                               & = 4 + 8 =  6 + 6  = 8 + 4 
                    \label{eqn:polab}
\end{align}
These options lead to $ a_{1,2} . b_1 = -1 , 0, 1 $ respectively. Let's say one of these inner products is non-zero; i.e., (without loss of generality) $ a_1 . b_1 = \pm 1 $. Subsequently, $ || a_1 \mp 2 b_1 ||^2 = 14 $, and must correspond to an (unsigned) spin-1 operator; i.e., $ \llangle a_1 \mp b_1 \rrangle ^2 = 2 \langle a_1 , b_1 \rangle = \pm 8 $ (see \eqref{eqn:Z3} and \eqref{eqn:newDLreln}).\footnote{Note that we've used $ \llangle a_1 \rrangle ^2 = \llangle b_1 \rrangle ^2 = 0 $ -- throughout this proof, we'll be making sure that all named vectors ($a, b, c$, etc) have zero Lorentzian norm; their linear combinations may not.} But we also have $ || a_1 \mp b_1 ||^2 = 4 $, which we know is a scalar: $ \llangle a_1 \mp b_1  \rrangle ^2 = 2 \langle a_{1} , b _1 \rangle = 0 $. We summarise these points in the following box:
\begin{align}
\boxed{
\begin{array}{rl}
    || a_1 \mp b_1 ||^2   = 4 & \Rightarrow \langle a_1 , b_1 \rangle = 0 \\
    || a_1 \mp 2 b_1 ||^2 = 14 & \Rightarrow \langle a_1 , b_1 \rangle \neq 0
\end{array}
}
\label{eqn:exContra}
\end{align}
We will use these boxed equations to denote possible contradictions. In the present case, we say that we will run into \eqref{eqn:exContra} if we choose $ a_1 . b_1 = \pm 1 $; and a similar contradiction if we go with $a_2 . b_1 = \pm 1 $. Hence, we are forced into the choice: $a_1 . b = a_2 . b = 0 $, in which case we can construct another contradiction:
\begin{align}
\boxed{
\begin{array}{rl}
    || a_1 + b_1 ||^2  = 6  & \Rightarrow \langle a_1 , b_1 \rangle = 0 \\
    || a_2 + b_1 ||^2   = 6 & \Rightarrow \langle a_2 , b_1 \rangle = 0 \\
    || a_1 + a_2 ||^2  = 4 & \Rightarrow \langle a_1 , a_2 \rangle = 0 \\ 
    ||| a_1 + 2a_2 + b_1 ||^2 = 14 & \Rightarrow 
            \langle a_1 , a_2 \rangle + \langle a_1 , b_1 \rangle + \langle a_2 , b_1 \rangle \neq 0 \\ 
\end{array}
}
\label{eqn:exContra2}
\end{align}
The third equation states that none of the operators with $ \Delta = \frac{7}{4} $ are scalars. In conclusion, all choices lead to contradictions and hence the assumption of a Narain progenitor must be false.

\begin{figure}

\begin{tikzpicture}[node distance=3cm]

    \node (start) [startstop] { \begin{tabular}{c}
                                     $ ||a_1||^2 = 2 , \quad ||a_2||^2 = 2 $  \\
                                     $ a_1 . a_2 = 0 $ \\
                                     $ ||b_1 ||^2 = 4 $
                                \end{tabular}
                                };
    \node (ab1) [contra, below of=start, xshift=-2cm, yshift=-1cm] {\eqref{eqn:exContra}};
    \node (ab2) [contra, below of=start, xshift=2cm, yshift=-1cm] {\eqref{eqn:exContra2}};

    \draw [arrow] (start) -- node[above, sloped] {$ a_1 . b_1 = \pm 1 $} (ab1);
    \draw [arrow] (start) -- node[below, sloped] {$ a_2 . b_1 = \pm 1 $} (ab1);
    \draw [arrow] (start) -- node[above, sloped] {$ a_{1,2} . b_1 = 0 $} (ab2);
    
\end{tikzpicture}

\caption{The outline of the argument in the case of $\tilde{Z}_{3}$ (and also $\tilde{Z}_{1}$, see appendix \ref{elementaryproof}). The colored boxes indicate contradictions. }

\label{fig:Z13}

\end{figure}
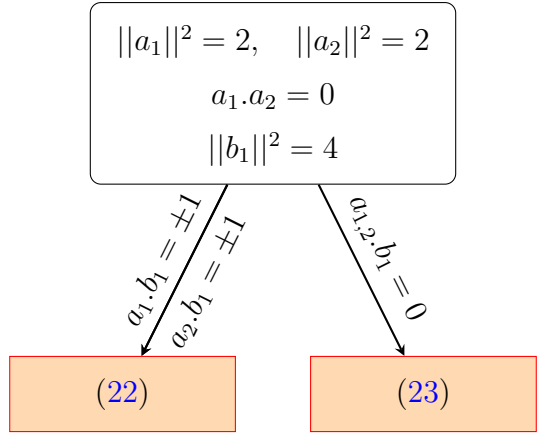

The proof above in terms of elementary operations  surprisingly generalizes to all six $Z_i$, although the number of steps could be significant, see Appendix \ref{elementaryproof}.

\subsection{Proof by constructing all lattices explicitly}
\label{alllattices}

In this section, we give  alternative and more general proof that none of the impostor functions $Z(\tau,\bar \tau)$ obtained from  by \eqref{eqn:fake} originate from CFTs of the Narain type. 

We saw in the previous section that would be Euclidean lattice $2\LL_3$ had to be even. This situation is completely general for all $Z_i$.
Consider any  ``partition function'' $\Z$ obtained via \eqref{eqn:recipe}. It  is a polynomial in
\begin{align*}
\theta_3 \bar{\theta}_3 + \theta _4 \bar{\theta}_4 
    & =
        1+4 \varrho +\varrho ^2 \left(2 t^2+\frac{2}{t^2}\right)+4 \nwDelta ^4+\nwDelta ^5 \left(4 \nwl^4+\frac{4}{\nwl^4}\right)+\dots 
\nonumber \\
\theta_3 \bar{\theta}_3 - \theta _4 \bar{\theta}_4 
    & =
       \frac{2 \sqrt{\nwDelta } (\nwl+1)}{\sqrt{\nwl}}+\frac{4 \nwDelta ^{5/2} \left(\nwl^3+1\right)}{\nwl^{3/2}}+\frac{2 \nwDelta ^{9/2} \left(\nwl^9+1\right)}{\nwl^{9/2}}+\dots
\nonumber \\
\theta_2 \bar{\theta}_2
    & =
        2 \sqrt[4]{\nwDelta }+\nwDelta ^{5/4} \left(2 \nwl+\frac{2}{\nwl}\right)+2 \nwDelta ^{9/4}+\nwDelta ^{13/4} \left(2 \nwl^3+\frac{2}{\nwl^3}\right)+\nwDelta ^{17/4} \left(2 \nwl^2+\frac{2}{\nwl^2}\right)+\dots
\end{align*}
and hence the series expansion of $\Z$ only contains positive integer powers of $ \varrho ^{1/4}$. Assuming that $\Z$ is genuine, meaning it is a Siegel theta function of some Narain lattice $\LL$, the sq-length of all vectors of $\LL$  are half-integers. Therefore, $ 2 \LL $ is an even lattice with respect to the Euclidean inner product and
in particular, all Euclidean norms are even integers.\footnote{Since $ \LL $ is already even and self-dual with respect to $\eta $; $ 2 \LL $ is also integral (but no longer self-dual) with respect to $\eta$. However, we will not use this fact directly.} 

Since $\LL$ is self-dual with respect to $\eta$, its determinant is $1$. Therefore $2\, \LL$, understood 
as a Euclidean lattice, is even and has determinant $2^{2n}$. All such lattices -- for small $n$ -- can be constructed explicitly and their Euclidean theta series can be compared with $\Z(\nwDelta^4,\nwDelta^4)$.  This prompts the  following strategy:  construct all $2n=6$-dimensional even lattices with determinant $2^{6}$ and compute their theta series up to some order. We will see that there are no lattice theta series matching $\Z_i(\nwDelta^4,\nwDelta^4)$ for $i=1,2$, thus ruling them out as torus partition functions of Narain theories. 
Yet we find lattices reproducing $\Z_i(\nwDelta^4,\nwDelta^4)$ for $i=3,4,5,6$, for which the argument has to be extended. Namely, for each Euclidean  lattice yielding a particular $\Z_i(\nwDelta^4,\nwDelta^4)$, we would like to equip it with a Lorentzian inner product $\eta$ such that Siegel theta function matches $\Z_i(\nwDelta^4 t^4,\nwDelta^4 t^{-4})$. This leads to a number of linear equations for $\eta$ written in some particular basis, which in all cases have no solutions. Here again, absence of zero spin states with $\Delta=7/4$ play a crucial role.

We will illustrate these points  below, using an $ n =1 $ enumerator polynomial as a working example. However,  in this case the corresponding $Z$  comes from an actual Narain CFT, and we will end up rediscovering this fact.

\subsubsection{Taxonomy of integral lattices\label{sec:classification}}
In \eqref{LambdaBg}, the lattice generating matrix was written in a particular form, e.g.~satisfying \eqref{eta}. More generally a Euclidean $2n$-dimensional lattice can be defined as an equivalence class of  lattice generating matrices $\Lambda$, where $\Lambda \sim \Lambda'$ if they are related by a rotation and renumeration of lattice points, 
\begin{equation}
\Lambda'=O\Lambda S,\qquad O\in O(2n,\mathbb{R}),\qquad S\in GL(2n,\mathbb{Z}).
\end{equation}
Alternatively, a lattice is an equivalence class of quadratic forms  $G=\Lambda^t \Lambda$ 
up to $S\in GL(2n,\mathbb{Z})$ transformations $G\sim S^T G S$.

To construct all even (and hence integral) Euclidean lattices in $\R^{2n}$ with the determinant $2^{2n}$ we use the following result (our statement is a fusion of two lemmas in \cite{10.2307/2398315}):
\begin{lemma}
Any integral Euclidean lattice of determinant $2\mathfrak{D}$ (with $\mathfrak{D}^2 \in \mathbbm{Z} $) can be obtained from an integral Euclidean lattice of determinant $\mathfrak{D}$ via 
\begin{equation}
\Lambda_{2\mathfrak{D} }=\Lambda_{\mathfrak{D}} B,
\end{equation}
where $B$ is some integer-valued matrix with ${\rm det}\, B=2$.
\end{lemma}
It is obvious that if $\Lambda_\mathfrak{D}$ generates an integral lattice of determinant $\mathfrak{D}$, then 
$\Lambda_\mathfrak{d} B$ generates an integral lattice of determinant $2\mathfrak{D}$. The non-trivial part here is that all integral lattices of determinant $2\mathfrak{D}$ can be obtained this way.

Let us now restrict our attention to the case at hand -- we have $ n = 3$. It follows from the above, that all integral lattices in $\mathbb{R}^6$ of determinant $2^6$ are given by 
\begin{equation}
\Lambda_{i_1\dots i_6}= B_{i_1}\dots B_{i_6}
\label{eqn:Bstring}
\end{equation}
where $B_i$ exhaust all possible integer-valued matrices of determinant 2. Here we used the fact that the only unimodular  (integral and of determinant one) lattice in $\R^6$ is $\mathbb{Z}^6$  and its generating matrix can be chosen to be the identity matrix.\footnote{This statement is true in any dimension less than $8$. In $\mathbb{R}^8$, besides $\mathbb{Z}^8$, there is also a unimodular root lattice $E_8$.}
Using the $GL(6,\mathbb{Z})$ action from the right we can bring all $B_{i_k}$ to Hermite normal form, namely lower triangular with non-negative integer matrix elements, and  the  diagonal  matrix elements (pivots) are largest in each row. There are $2^6-1$ such matrices. Using permutations of rows, which is an element of $O(6,\mathbb{R})$ one can bring $B_{i_1}$ to one of $6$ possible forms (last pivot is two, all others are one, there are $0\leq k\leq 5$ ones in the last row located arbitrarily), hence bringing the total number of combinations  in \eqref{eqn:Bstring} to $6(2^6-1)^5$. We should stress that not all combinations yield distinct lattices, yet possible equivalence relations between different $B_{i_1}\dots B_{i_6}$ combinations are difficult to identify. We illustrate the construction with a 2d example below. 

\subsubsection*{All integral determinant-$4$ lattices in  $\R^2$}
To illustrate our method, we compute all two-dimensional integral lattices of  determinant $4$. There are $2^2-1=3$ possible matrices $B_i$ in the Hermite form
	\begin{align}
D_{1,1}  = \left(\begin{array}{cc}
2 & 0\\
0 & 1
\end{array}\right),\quad 
	D_{1,2}  = \left(\begin{array}{cc}
1 & 0\\
0 & 2
\end{array}\right),\quad 
	D_{2,2}  = \left(
\begin{array}{cc}
 1 & 0 \\
 1 & 2 \\
\end{array}
\right).
	\end{align}
Naively, there are $9$ different combinations. Using permutations of rows we can reduce $B_{i_1}$ to be either $D_{1,2}$ or $D_{2,2}$.
There are then 6 possible combinations, which  generate 4 distinct lattices/quadratic forms/theta series. 

As explained before, this list must contain the (appropriately rescaled) lattices of all $ c= 1 $ Narain theories  with vertex operators of quarter-integer dimensions. There are three such theories -- the compact scalars with $ R = 1 , 2 , \sqrt{2} $.  The first two are code theories, and being T-dual to each other, their lattices coincide. The generating matrices for the corresponding Narain lattices are:
\begin{align}
2 \Lambda _{R = 1}  = \sqrt{2}  \begin{pmatrix}
 								1 & 0 \\
 								0 & 2
 								\end{pmatrix}
 						\sim
 						D_{2,2} D_{1,2} \sim D_{2,2} D_{2,2},
 	\nonumber \\
   2 \Lambda _{R = \sqrt{2}}  =   \begin{pmatrix}
 								2 & 0 \\
 								0 & 2
 								\end{pmatrix}
 						\sim
	 						 D_{2,2} D_{1,1}\sim D_{1,2} D_{1,1}.
 \end{align} 
In the two-dimensional case, the equivalence of lattice-generating matrices is easy to establish explicitly, with a more practical approach would be first to compare corresponding theta series calculated to a sufficient order.  

The other two combinations $D_{1,2} D_{1,2}$ and $D_{1,2} D_{2,2}$ do not generate even integral lattices, and hence cannot correspond to Narain theories  with quarter-integer spectra. 

\subsubsection*{The full list of $d=6$ theta series and the result of Euclidean analysis \label{sec:genexplicit}}

A naive attempt to write down all  $6(2^6-1)^5$ combinations $B_{i_1}\dots B_{i_6}$ is challenging even using computer algebra. In fact, we do not need the lattices themselves but only their theta series, to compare with $\Z_i(\nwDelta^4,\nwDelta^4)$.
There is a very useful trick\footnote{We thank Ohad Mamroud for a valuable discussion on this point.},  described in Appendix \ref{trick}, which allows for efficient construction of the Siegel theta series of $B_{i_1}\dots B_{i_{r-1}}B_{i_r}$ from the generating matrix of
$B_{i_1}\dots B_{i_{r-1}}$, thus reducing the number of necessary matrix multiplications to $6(2^6-1)^4$. While this is still a large number, this task can be performed on a computer cluster, yielding no combinations with theta series matching $\Z_1$ and $\Z_2$ and many different combinations matching $\Z_i$ for 
 $i=3 - 6$.  These combinations will be used for further analysis in Section \ref{sec:lor}. 
We also conclude  at this point  that the partition functions $Z_{1,2}(q, \bar{q})$ cannot come from any Narain CFTs.

We close this section with a trick that could help to drastically reduce the computation time. After each $B_{i_k}$ is bought to the Hermite normal form, the matrix product  \eqref{eqn:Bstring} is of the form:
\begin{eqnarray}
D=\left( 
\begin{array}{cccc}
2^{p_1} & 0 & 0 &\dots\\
* & 2^{p_2} & 0&\dots\\
* & * & 2^{p_3}&\dots\\
\dots &\dots&\dots&\dots
\end{array}
\right)  \label{HNF}
\end{eqnarray}
The sum $p_1+\dots +p_d=r$ is fixed by the value of the determinant. We are  interested in the case $r =d= 6$. To simplify things further, we use the equivalence relation -- the row permutations acting from the left  to satisfy an additional condition  $p_1\leq p_2\leq \dots \leq p_d$.\footnote{We do not prove that this is always possible, but checked it explicitly for all cases with $r=d\leq 5$.\label{subtlety}} The diagonal of such matrices are labelled  by Young diagrams with $r$ boxes.

The lattices we are interested in are not merely integral but even. Hence we can demand that the squared-sum of elements within each row of $D$ be even. Finally, using reflections along each of the $d$ axes (which are lattice equivalences) we can make sure that in each column, moving downwards starting from below the diagonal, the first non-zero element must be less or equal than the corresponding pivot divided by two. So if the first non-zero element in the column numbered $d-1$ is $D_{d,d-1}$, we must have $D_{d,d-1} \leq 2^{p_d-1}$. 

With these ``gauge fixing'' conditions, in the case of $d=r=6$  there are only $7,725,064$ distinct matrices $D$. All of them can be easily  generated on a laptop in a matter of minutes. A brute force calculation of the corresponding  theta series then shows that none of them match $ \Z_{1}(\varrho ^4,\varrho ^4)$ or $\Z_2(\varrho ^4,\varrho ^4)$, in agreement with the discussion above. 
At first glance, one finds many  thousands of  $D$ that yield $\Z_i(\varrho ^4,\varrho ^4)$ for each $i=3-6$, but upon further analysis they all turn out to be equivalent. In conclusion, for each $ i =3- 6 $, we find exactly one lattice with the theta series matching $\Z_i(\varrho ^4,\varrho ^4)$-- with the generators given by:
\begin{eqnarray}
\label{L34}
D_3=
\left(
\begin{array}{cccccc}
 1 & 0 & 0 & 0 & 0 & 0 \\
 0 & 1 & 0 & 0 & 0 & 0 \\
 0 & 0 & 2 & 0 & 0 & 0 \\
 0 & 0 & 0 & 2 & 0 & 0 \\
 0 & 1 & 0 & 0 & 4 & 0 \\
 1 & 0 & 0 & 0 & 0 & 4 \\
\end{array}
\right)
,\qquad 
D_4=
\left(
\begin{array}{cccccc}
 1 & 0 & 0 & 0 & 0 & 0 \\
 0 & 1 & 0 & 0 & 0 & 0 \\
 0 & 0 & 2 & 0 & 0 & 0 \\
 0 & 1 & 0 & 2 & 0 & 0 \\
 1 & 0 & 0 & 2 & 4 & 0 \\
 2 & 2 & 2 & 0 & 0 & 4 \\
\end{array}
\right)
,\\
\label{L56}
D_5=
\left(
\begin{array}{cccccc}
 1 & 0 & 0 & 0 & 0 & 0 \\
 0 & 1 & 0 & 0 & 0 & 0 \\
 0 & 1 & 2 & 0 & 0 & 0 \\
 0 & 1 & 0 & 2 & 0 & 0 \\
 0 & 1 & 0 & 0 & 4 & 0 \\
 1 & 2 & 0 & 0 & 0 & 4 \\
\end{array}
\right)
,\qquad 
D_6=
\left(
\begin{array}{cccccc}
 1 & 0 & 0 & 0 & 0 & 0 \\
 0 & 1 & 0 & 0 & 0 & 0 \\
 0 & 1 & 2 & 0 & 0 & 0 \\
 0 & 1 & 0 & 2 & 0 & 0 \\
 1 & 2 & 0 & 0 & 4 & 0 \\
 2 & 1 & 0 & 0 & 0 & 4 \\
\end{array}
\right),
\label{eqn:latticeList}
\end{eqnarray}
The corresponding theta series read:
\begin{align*}
\Theta_3 & = Z_3 ( \varrho ^4 , \varrho ^4 ) = 1+4 \varrho+8 \varrho^2+16 \varrho^3+28 \varrho^4+40 \varrho^5+64 \varrho^6+96 \varrho^7+124 \varrho^8+\dots,\\
\Theta_4 & = Z_4 ( \varrho ^4 , \varrho ^4 ) = 1 + 8 \varrho^2 + 16 \varrho^3 + 28 \varrho^4 + 64 \varrho^5 + 64 \varrho^6 + 96 \varrho^7 + 124 \varrho^8+\dots,\\
\Theta_5 & = Z_5 ( \varrho ^4 , \varrho ^4 ) = 1 + 2 \varrho + 4 \varrho^2 + 24 \varrho^3 + 44 \varrho^4 + 20 \varrho^5 + 32 \varrho^6 + 144 \varrho^7 + 
 188 q^8+\dots,\\
\Theta_6 & = Z_6 ( \varrho ^4 , \varrho ^4 ) = 1 + 4 \varrho^2 + 24 \varrho^3 + 44 \varrho^4 + 32 \varrho^5 + 32 \varrho^6 + 144 \varrho^7 + 188 \varrho^8+\dots
\end{align*}

\subsubsection{The Lorentzian step \label{sec:lor}}

Now we proceed to prove that $Z_i \lb q , \bar q \rb $ for $ i = 3- 6 $ cannot arise from any Narain CFT. That is for $ i = 3-6$, we show that no lattice which yields the theta series $  \Theta_{\Lambda/2}(\varrho) = \tilde{Z}_i \lb \varrho , \varrho \rb$ also satisfies $  \Theta_{\Lambda/2, \eta} \lb q, \bar q \rb = \tilde{Z}_i \lb q , \bar q \rb$ at the level of the Siegel theta series, with some Lorentzian inner product $\eta$. 

We start with some particular $Z_i$; for example, with $ i = 3 $. As we explained above, we have  constructed an exhaustive list of candidate lattice-generating matrices such that corresponding theta series satisfy
$\Theta_\Lambda(\varrho)=\Z_3(\varrho^4,\varrho^4)$. While we expect that all these matrices are equivalent, i.e.~correspond to the same Euclidean lattice, we can not rigorously prove it (see the footnote \ref{subtlety}) and deal with them one by one.  We would like to know if there is an appropriate $\eta$, a symmetric $6 \times 6$ matrix with the signature $(+,+,+,-,-,-)$ such that 
\begin{align}
\Z_3 (\varrho ^4 t , \varrho ^4 t^{-1} )
	& 
	= \sum _{\substack{v = \Lambda  m \\ m \in \mathbbm{Z}^{2n} }} \varrho ^{ \frac{1}{2} v^t v }\,  t^{\frac{1}{8} v^t \eta v}
	\nonumber \\
	& = 
		1 + 4 \varrho  + 8 \varrho ^2 + 16 \varrho ^3 + 4 \lb 5 + t+t^{-1} \rb \varrho ^4 + 
 4 \lb 8 + t+t^{-1} \rb \varrho ^5 
 \nonumber
 \\
 & \qquad \qquad 
 + 16 \lb 2 + t+t^{-1} \rb \varrho ^6  + 48 \lb t+t^{-1} \rb \varrho ^7+\dots   
\end{align}
We start by noting that all lattice vectors  $v$ that have sq-length $2, 4, 6$ have zero norm under $\eta$; and those with sq-length $14$ have $\eta$ norm-squared $\pm 8$. For each $\Lambda$ we can easily  identify all the short vectors explicitly, leading to a system of linear (non)equalities:
\begin{align}
v ^t\, \eta\, v
 = 0 ,
 & \qquad \text{for} \ || v||^2 = 2, 4, 6
 \nonumber \\
v ^t \eta v
 \neq 0 ,
 & \qquad \text{for} \ || v ||^2 = 14.
\label{system}
\end{align}
However, in all cases, this system has no solutions, which we checked on a cluster using the list of all suitable $\Lambda=B_{i_1}\dots B_{i_6}$. 

The same applies to all the other cases: 
\begin{align*}\nonumber
\Z_4(\varrho ^4 t, \varrho ^4t^{-1}) & =1 + 8 \varrho ^2 + 16 \varrho ^3 + 4 (5 + t+t^{-1}) \varrho ^4 + 16 (2 + t+t^{-1}) \varrho ^5 
\\
 & \qquad \qquad + 16 (2 + t+t^{-1}) \varrho ^6 + 48 (t+t^{-1}) \varrho ^7+\dots
\\
\Z_5(\varrho ^4 t, \varrho ^4t^{-1})
& =
1 + 2 \varrho  + 4 \varrho ^2 + 24 \varrho ^3 + (28 + 8 t+8t^{-1}) \varrho ^4 + 
 2 (8 +  t+t^{-1}) \varrho ^5 
\\
 & \qquad \qquad 
 + 8 (2 + t+t^{-1}) \varrho ^6 + 72 ( t+t^{-1}) \varrho ^7 +\dots 
\\
\Z_6(\varrho ^4 t, \varrho ^4t^{-1})
& =
1 + 4 \varrho ^2 + 24 \varrho ^3 + (28 + 8t+8t^{-1}) \varrho ^4 + 8 (2 + t+t^{-1}) \varrho ^5 
\\
 & \qquad \qquad
+  8 (2 + t+t^{-1}) \varrho ^6 + 72 (t+t^{-1}) \varrho ^7+\dots 
\end{align*}  
In each case the $\Delta=7/4$ states have non-zero spins, while states of dimension $2,4,6$ (if any) are scalars; and in all cases, meaning for all suitable lattices $\Lambda=B_{i_1}\dots B_{i_6}$ the system \eqref{system} has no solutions. 

For the four representatives 
\eqref{L34},~\eqref{eqn:latticeList}; we can see explicitly why this happens: some (but not all) vectors $v_{14}$ of sq-length $14$ can be expressed through vectors of sq-length $2,4,6$ in the sense:
\begin{align}
v_{14} v^t_{14}=\sum_i a_i v_i v_i^t,
\end{align}
with some coefficients $a_i$, 
and $i$ running through all {\it scalar} vectors of sq-length less or equal to $6$. This immediately rules out $v_{14}^t \eta v_{14}\neq 0$. This concludes the proof that none of $Z_i$'s is a torus partition function of some Narain theory.  

In conclusion, it is interesting to note that the elementary proof of Section \ref{sec:elementary} and the more general proof of this section both rely on ``representing'' $\Delta=7/4,\ell=\pm 1$ vectors starting from the shorter scalar  ones.

\subsubsection*{Example: the $ c = 1 $ case\label{sec:c=1}}

Let's illustrate the procedure in the $c =1 $ case --  for the enumerator polynomial $ W(x,y, z) = x + z $. The vertex partition function in this case is given by:
 \begin{align}
\tilde{Z} _W \lb \varrho ^4 t , \varrho ^4 t ^{-1} \rb 
		& =
		1+2 \varrho +4 \varrho ^4+2 \varrho ^5 \left( t + t^{-1} \right)+2 \varrho ^8 \left(t^2 + t^{-2}\right)+2 \varrho ^9+ 2 \varrho ^{13} \left( t^3 + t^{-3} \right)+ \dots
	 \label{eqn:ZR1full}
\end{align}
This is the partition function of the $R=1$  Narain theory, but  this is not known to us yet. In  Section \ref{sec:classification}, we  identified a Euclidean  lattice with the theta function $\tilde{Z} _W$,
\begin{align}
\tilde{Z} _W \lb \varrho ^4, \varrho ^4  \rb 
	& = 
		\sum _{ r \in \mathbbm{Z}^{2} } \varrho ^{ \frac{1}{2} r^t  \Lambda ^t   \Lambda  r },\qquad \Lambda=D_{2,2}D_{1,2}.
\end{align}
We would like to find a Lorentzian inner product matrix $\eta$ such that 
\begin{align}
\tilde{Z} _W \lb \varrho t, \varrho t^{-1}  \rb 
	& = 
		\sum _{ r \in \mathbbm{Z}^{2} } \varrho ^{ \frac{1}{8} r^t  \Lambda ^t   \Lambda  r } t^{ \frac{1}{8} r^t  \Lambda ^t \eta \Lambda  r },
\label{match}
\end{align}
or show that it does not exist. 
First, we identify all vectors of sq-length two, $v_2=\pm (1,1)$ and four $v_4=(\pm 2,\pm 2)$. Demanding that these vectors be scalars restricts $\eta$ to be of the form 
\begin{equation}
\eta=\begin{pmatrix}
			\eta_{11} & 0 \\
			0 & - \eta_{11}
		\end{pmatrix}. \label{etac}
\end{equation}
Then we consider a vector of sq-length five, $v_5=(-3,1)$. Demanding that this vector has (unsigned) spin one; i.e., $\tfrac{1}{8} v_5^t \eta v_5=1$, fixes $\eta_{11}=1$. This fixes $\eta$ and one can check, at least pertubatively,  that \eqref{match} is satisfied. Clearly \eqref{etac} is related to \eqref{eta} by an orthogonal $O(2,\R)$ transformation, which finishes the proof -- starting from the polynomial $W$, we have constructed a Narain lattice (of the $c=1$ compact scalar at radius $R=1$) that yields the same Siegel theta series as $\Z_W$.

\section{Excluding unitary CFTs \label{sec:final}}

In this section, we show that $Z_{3-6}$ cannot come from any unitary CFT with a local stress tensor. In particular, this also proves that they cannot come from the Narain CFTs and constitutes an independent proof of this claim. We do this by first recalling that the algebra formed by spin-1 currents in  a unitary CFT with a local stress tensor must be a direct sum of affine Lie algebras and Heisenberg algebras\footnote{By which we mean the chiral algebra of a non-compact free boson.}. Then we show that the spectra of $Z_{3-6}$ cannot be organised into representations of any such algebra. Our analysis will prove inconclusive for $Z_{1,2} $.

Consider the full set of holomorphic, spin-1  operators in a unitary CFT with a unique ground state. Since we are dealing with a unitary theory, we can assume that this set has a basis of Hermitian operators -- say $j^a (z)$ -- orthonormal under the two-point function:
\begin{align}
    \langle j^a (z) j^b (0) \rangle = \frac{1}{z^2} \delta ^{a b}
\end{align}
The products of these operators have to be of the form:
\begin{align}
    j^a (z) j^b (0) \sim \frac{1}{z^2} \delta ^{a b} + \frac{1}{z} \iu \tilde{f}^{ab}_{c} j^c (0)  
\end{align}
It then follows from crossing symmetry that the structure constants $\tilde{f}^{ab}_{c}$ satisfy the Jacobi-Lie identity. This means that the zero modes of $j^a(z)$ form a (real) Lie algebra $\mathfrak{g}$. The two-point function is a positive-definite bi-invariant (i.e., to both left and right action) metric on this Lie algebra, and the three-point functions $ \tilde{f}^{abc} $ are completely antisymmetric thanks to Bose symmetry. A positive-definite metric and completely antisymmetric structure constants mean that  $\mathfrak{g}$ is constrained to be a direct sum of simple algebras and $\mathfrak{u}(1)$'s, see for instance \cite{MILNOR1976293}. Subsequently, the $j^a(z)$ form a direct sum of affine Lie algebras and Heisenberg algebras.

In what follows, we will constrain the list of algebras that can constitute $\mathfrak{g}$ for each of the $\tilde{Z}_i$'s and for $ i = 3-6$ conclude that this list is empty. Our tools are going to be  crossing symmetry (i.e., the conformal bootstrap) and some representation theory. For $Z_{1,2}$ we will find that the spectrum falls into representation of $\lb \mathfrak{u}(1)_4 \rb ^3$ chiral algebra, thus leading to no apparent contradiction. This does not necessarily mean $Z_{1,2}$ are genuine, but this is a possibility. To illustrate this point, toward the end of this section, we consider a $c=2$ code theory with the spectrum falling into representations of $ \mathfrak{su}(2)_1 \oplus \mathfrak{su}(2)_1$.

\subsection{Polynomials 5 and 6}
Since $\mathfrak{g}= \bigoplus _{i=1}^s \mathfrak{g}_i $ is a direct sum of simple and Abelian Lie algebras, we will rescale the $j^a$'s from each individual simple $\mathfrak{g}_i$ (we keep the Abelian generators as they are) to their conventional normalisation:
\begin{align}
    \mathfrak{k} ^{ab} = f^{ap}_q f^{bq}_p = 2 g_i ^{\vee}\delta^{ab},\ \text{for }a,b \in \mathfrak{g}_i
\end{align}
where $ g_i ^{\vee} $ is the dual Coxeter number of $\mathfrak{g}_i$ (see Table \ref{tab:simpleAlg}). The OPE is then:
\begin{align}
    j^a (z) j^b (0) \sim \frac{1}{z^2}\kappa ^{ab}  + \frac{1}{z} \iu f^{ab}_{c} j^c (0)  
    \label{eqn:curralg}
\end{align}
$\kappa ^{ab} $ here is a direct sum of matrices of the form $ k_i \mathbbm{1}$ that tell us the levels of representations of each of the simple factors ($k = 1$ for the Abelian factors as per our normalisation). The levels of the simple factors are positive integers thanks to unitarity.
Moreover, the stress-energy tensor breaks up into two commuting factors:
\begin{align}
T & = T_{\text{sug}} + T_{\text{rest}}
\end{align}
where the $T_{\text{sug}}$ is given by the Sugawara construction. This also means that (again, by unitarity):
\begin{align}
c & = c_{\text{rest}} + \sum_{i=1}^s c_i 
 \geq \sum_{i=1}^s c_i
    \nonumber \\
c_i & = c_{sug} (\mathfrak{g}_i, k_i) = \frac{ k_i \text{ dim} \ \mathfrak{g}_i }{ k_i + g_i^{\vee} }
\end{align}
Now consider: 
\begin{align}
Z_5 \lb \tau , \bar{ \tau } \rb
	& = 
		\frac{1}{ \lb \eta \lb q \rb \bar{\eta} \lb \bar{q } \rb  \rb ^3 } \lb 1 + 2 \lb q \bar{q} \rb ^{\frac{1}{8}} + 4 \lb q \bar{q} \rb ^{\frac{1}{4}} +  24 \lb q \bar{q} \rb ^{\frac{3}{8}} +  8 q + 8 \bar{q}  + ...    \rb
\end{align}
We can see that there are eight extra conserved currents, on top of $3$ ($+\ \bar{3}$) that come from the $\eta $ terms in the denominator. Together, they must give rise to an affine algebra $ \mathfrak{g}_k $ with:
\begin{align}
\sum_{i=1}^s \frac{ k_i \text{ dim} \ \mathfrak{g}_i }{ k_i + g_i^{\vee} } \leq 3
\label{eqn:csug}
\end{align}
and dim $\mathfrak{g} = 11$. We will now use the inequality \eqref{eqn:csug} to narrow down  possible options for $\mathfrak{g}_i$.
\begin{table}\centering
\ra{1.5}
\begin{tabular}{@{}cccc@{}}\toprule
$\mathfrak{g}$ & dim $\mathfrak{g}$ & rank $\mathfrak{g}$ & $g^{\vee}$    \\
\cmidrule{1-4}
$\mathfrak{u}(1)$ & $ 1 $ & $ 1 $ & $ 0 $ \\ 
 \hline
 $\mathfrak{su}(N)$ & $ N^2 - 1 $ & $ N - 1$ & $ N $ \\ 
 \hline
 $\mathfrak{so}(2N + 1 )$ & $2N^2 + N$ & $N$ & $ 2N - 1$ \\
 \hline
 $\mathfrak{sp}(N )$ & $2 N^2 + N$ & $N$ & $N + 1$ \\
 \hline
 $\mathfrak{so}(2N )$ & $2 N^2 - N$ & $N$ & $2 N -  2$ \\
 \hline
 $E_6 $ & $ 78 $ & $6 $ & $ 12 $ \\
 \hline
 $E_7 $ & $ 133 $ & $ 7 $ & $ 18 $ \\
 \hline
 $E_8 $ & $ 248 $ & $ 8 $ & $ 30 $ \\
 \hline
 $F_4 $ & $ 52 $ & $ 4 $ & $ 9 $ \\
 \hline
 $G_2 $ & $ 14 $ & $ 2 $ & $ 4 $ \\
 \hline
\bottomrule
\end{tabular}
\caption{A list of various simple algebras and their properties (borrowed from \cite{Fuchs:1997jv}).}
\label{tab:simpleAlg}
\end{table}

For each of the factors:
\begin{align}
 \text{rank } \mathfrak{g}_i & \leq c _i \leq  \text{dim }\mathfrak{g}_i \leq 11
 \end{align} 
A quick look at Table \ref{tab:simpleAlg} then tells us that each simple factor must be one of the following: $\mathfrak{su}(2) \simeq \mathfrak{sp}(1)\simeq \mathfrak{so}(3) $, $\mathfrak{su}(3) $, $\mathfrak{sp}(2) \simeq \mathfrak{so}(5) $, and $\mathfrak{so}(4) $. We haven't constrained  possible  values of levels yet, and list corresponding central charges as a function of $k$ in Table \ref{tab:levels}.

\begin{table}[h!]\centering
\ra{1.5}
\begin{tabular}{@{}ccc@{}}\toprule
$\mathfrak{g}$ & dim $\mathfrak{g}$ & $c_{sug} (\mathfrak{g}, k ) $    \\
\cmidrule{1-3}
 $\mathfrak{u}(1)$ & $ 1 $ & $ 1 $\\ 
 \hline   
 $\mathfrak{su}(2)_k $ & $ 3 $ &  $ \frac{3k}{k+2} $ \\ 
 \hline
 $\mathfrak{su}(3 )_k $ & $ 8 $ &  $ \frac{8k }{k+3} $ \\
 \hline
  $\mathfrak{so}(4)_k $ & $ 6 $ &  $ \frac{6k }{k+2} $ \\
 \hline
 $\mathfrak{so}(5 )_k $ & $ 10 $ &  $ \frac{10k }{k+3} $ \\
\bottomrule
\end{tabular}
\caption{The central charges of the relevant simple algebras (and $\mathfrak{u}(1)$) as a function the level.}
\label{tab:levels}
\end{table}

We must now find a combination with eleven generators and total central charge $\leq 3$ -- the only possible choice (up to isomorphisms) is $ \mathfrak{su}(3 )_1 \oplus \mathfrak{su}(2 )_1 $. It also saturates the central charge bound and gives 
\begin{align}
T & = T_{\text{sug}}
\end{align}
 This means that the group action (i.e., the global charges) completely fixes the conformal dimensions. Also, the partition function must be of the form:
 \begin{align}
 Z_{5} \lb \tau , \bar{ \tau } \rb 
 	& = 
 		\left| \chi ^{ \mathfrak{su}(3 )_1 } _0 \chi ^{ \mathfrak{su}(2 )_1 } _0 \right| ^2  + ...
 \end{align}
 These are  spin-1 currents and their descendants, 
however,
\begin{align}
\chi ^{ \mathfrak{su}(3 )_1 } _0 \chi ^{ \mathfrak{su}(2 )_1 } _0
	& = 
		\frac{1}{ \lb \eta \lb q \rb \rb ^3 } \lb 1+ 8 q + 12 q^2 + ... \rb,
	\nonumber \\
Z_{5} \lb \tau , -\iu \infty \rb 
	& = 
            \frac{1}{ \lb \eta \lb q \rb \rb ^3 } 
		\lb 1 + 8 q + 6 q^2 + ... \rb,
\end{align}
meaning that there are not enough chiral operators in $Z_5 $ to fill out  relevant representation of $ \mathfrak{su}(3 )_1 \oplus \mathfrak{su}(2 )_1 $. We thus have our contradiction.

The chiral spectrum of $Z_6$ is the same as that of $Z_5$ and hence the argument to rule it out is the same.
 
\subsection{Polynomials 3 and 4}

We start by writing the impostor partition functions:
\begin{align}
Z_3 \lb \tau , \bar{ \tau } \rb
	& = 
		\frac{1}{ \lb \eta \lb q \rb \bar{\eta} \lb \bar{q } \rb  \rb ^3 } \lb 1 + 4 \lb q \bar{q} \rb ^{\frac{1}{8}} + 8 \lb q \bar{q} \rb ^{\frac{1}{4}} +  16 \lb q \bar{q} \rb ^{\frac{3}{8}} +  4 q + 4 \bar{q}  + ...    \rb
	\nonumber \\
Z_4 \lb \tau , \bar{ \tau } \rb
	& = 
		\frac{1}{ \lb \eta \lb q \rb \bar{\eta} \lb \bar{q } \rb  \rb ^3 } \lb 1 + 8 \lb q \bar{q} \rb ^{\frac{1}{4}} +  16 q^{\frac{9}{8}} \bar{q}^{\frac{1}{8}} + 16 q^{\frac{1}{8}} \bar{q}^{\frac{9}{8}} + 4 q + 4 \bar{q}  + ...    \rb
	\label{eqn:Z34}
\end{align}
It may be verified that the chiral spectra of these functions are identical. In both cases, there are seven spin-$1$ currents, and we can use the arguments of the previous subsection to conclude that the relevant Kac-Moody algebra is either $  \mathfrak{su}(2)_1 \oplus \mathfrak{su}(2)_1 \oplus \mathfrak{u}(1) $ or $\mathfrak{so}(4)_1 \oplus \mathfrak{u}(1) $ -- we can restrict our attention to the first case. This time though, the chiral spectrum has more than enough states to support the relevant representation of algebra:
\begin{align}
    Z_{3,4} \lb \tau , -\iu \infty \rb - \frac{1}{ \eta \lb q \rb  } \lb \chi ^{ \mathfrak{su}(2 )_1 } _0 \rb ^2  = \frac{1}{ \eta \lb q \rb  } \lb 2 q^2 
+ 2q^8 + 2 q^{18} + ... \rb \lb \chi ^{ \mathfrak{su}(2 )_1 } _0 \rb ^2.
\label{eqn:extraXZ3}
\end{align}
As in the previous section, $ c_{sug} = c =3$ and we get
\begin{align}
T & = T_{\text{sug}}.
\end{align}
This means that the conformal dimensions are fixed by the global symmetry charges. The representations of $  \mathfrak{su}(2) \oplus \mathfrak{su}(2) \oplus \mathfrak{u}(1) $ are characterized by two spins and a charge, and we have:
\begin{align}
h_{(l_1 , l_2, \mathcal{Q} )} 
	& = 
		\frac{1}{3}\lb l_1(l_1+1) + l_2(l_2+1) \rb + 2 \gamma \mathcal{Q}^2
		\label{eqn:qh}
\end{align}
where $ \gamma $ is defined by $ T_{\text{sug}}^{ \mathfrak{u}(1)} = 2 \gamma \normord{j j} $, and cancels out the ambiguity from the normalisation of the $\mathfrak{u}(1) $ charge. We set $ \g \rightarrow 1 $ and normalise the charges accordingly.



We now  return to the non-chiral part of the spectrum (see \eqref{eqn:Z34}). For $Z_3$, this is led by an operator with dimensions $\lb \frac{1}{8} , \frac{1}{8} \rb $. The relevant representation of the $ \mathfrak{su}(2 )_1 \oplus \mathfrak{su}(2 )_1 \oplus \mathfrak{u}(1 ) $ has $ \lb l_1 , l_2 , \mathcal{Q} \rb =  \lb \overline{l}_1, \overline{l}_2, \overline{\mathcal{Q}} \rb = \lb 0, 0, \frac{1}{4} \rb $. It is easy to verify that $Z_3$ doesn't have enough states to fill in this representation.


The argument in the case of $Z_4$ is similar. The massive vector with dimensions $ \lb  \tfrac{9}{8}, \tfrac{1}{8} \rb$ has the lowest right-spin; hence, this operator must be a primary. This state can occur at the head of representations with $ \lb \overline{l}_1, \overline{l}_2, \overline{\mathcal{Q}} \rb = \lb 0, 0, \tfrac{1}{8} \rb $ and $ \lb l_1 , l_2 , \mathcal{Q} \rb = \lb 0, 0, \tfrac{3}{4} \rb $, $\lb 0, \tfrac{1}{2}, \tfrac{\sqrt{7}}{4} \rb $, or $\lb \tfrac{1}{2}, \tfrac{1}{2}, \tfrac{\sqrt{5}}{4} \rb $. It is easy to verify that $Z_4 $ doesn't have enough states to fill out any of these representations.

\subsection{Polynomials 1 and 2 } 

$Z_{1,2}$ come with extra spin-2 currents. The chiral spectrum is the same as the higher-spin algebra $\lb \mathfrak{u}(1)_4 \rb ^3$. But this time, we find that the rest of the spectrum also falls neatly into $\lb \mathfrak{u}(1)_4 \rb ^3$ representations: 
\begin{align}
Z_1 & = 
	\lb \left| \chi _0  \right| ^2 + \left| \chi _2 \right| ^2 + 2 \left| \chi _{1}  \right| ^2  \rb ^3 - 2 \left| \chi _{1}  \right| ^2 \lb \chi _0 ^2 - \chi _2 ^2 \rb \overline{ \lb \chi _0 ^2 - \chi _2 ^2 \rb }  
	\nonumber \\
Z_2 & = 
	\lb \left| \chi _0  \right| ^2 + \left| \chi _2  \right| ^2 + 2 \left| \chi _{1}  \right| ^2  \rb ^3 - 4 \left| \chi _{1}  \right| ^2 \lb \chi _0 ^2 - \chi _2 ^2 \rb \overline{ \lb \chi _0 ^2 - \chi _2 ^2 \rb }    
 \label{eqn:Z1exp}
\end{align}
where
\begin{align}
    \chi _t (\tau) & = \frac{1}{ \eta (\tau) } \sum_{r \in \mathbbm{Z}}  q^{2 \lb r + \tfrac{t}{4} \rb ^2 }
\end{align}
are $\mathfrak{u}(1)_4$ characters, and are related to the Jacobi theta functions:
\begin{align}
    \chi_0 (q) & = \frac{1}{2  \eta (\tau) } ( \theta_3( q ) + \theta_4( q ) ) \nonumber \\
    \chi_1 (q) = \chi_3 (q) & = \frac{1}{2 \eta (\tau) } \theta_2( q ) \nonumber \\
    \chi_2 (q) & = \frac{1}{2 \eta (\tau) } ( \theta_3( q ) - \theta_4( q ) ) 
\end{align}
Note that expanding the brackets in \eqref{eqn:Z1exp} produces a sum of  $\mathfrak{u}(1)_4$ characters with positive-integer coefficients. Hence, the methods from before do not work in this case and our analysis is inconclusive. But the results of the previous section mean that if $Z_{1,2}$ are indeed CFTs with the $(\mathfrak{u}(1)_4)^3 $ chiral algebra, they must be non-Narain theories. This would be interesting since it is widely assumed that compact CFTs with the $\mathfrak{u}(1)^n \times \mathfrak{u}(1)^n $ symmetry are Narain.

 \subsection{A genuine CFT example}

Finally, we consider an actual $c=2$ code CFT to illustrate how it passes the consistency checks discussed in this section.
Using the construction of \cite{ds}, up to T-duality there is a unique indecomposable $c=2$ code with the polynomial:
\begin{align}
    W & = x^2 + y^2 + 2z^2 \nonumber
\end{align}
leading to the code CFT with the partition function:
\begin{align}
    \tilde Z & = 1 + 4 q + 4 \bar{q} + 8 (q  \bar{q} ) ^{\tfrac{1}{4}} + 16 ( q \bar{q} ) ^{\tfrac{1}{2}} + ... 
\end{align}
Following the line of arguments outlined previously, the chiral algebra must have $ 6 $ holomorphic spin-$1$ currents with $ c \leq 2 $ -- we have a perfect fit with $\mathfrak{su}(2)_1 \oplus \mathfrak{su}(2)_1 $, with the rest of the spectrum also fitting into representations of this algebra:
\begin{align}
    \tilde{Z} = \lb \left| \chi ^{\mathfrak{su}(2)_1}_0 \right| ^2 + \left| \chi ^{\mathfrak{su}(2)_1} _{1/2} \right| ^2 \rb ^2
\end{align}
where $ \chi ^{\mathfrak{su}(2)_1}_{0,1/2} $ are the characters of the $ \mathfrak{su}(2)_1 $ representations with highest-weight states of spins $0$ and $1/2 $ respectively. Now, we see this is the partition function of two $ R = \sqrt{2} $ scalars -- even though the code is indecomposable, the CFT factors (see \cite{ds}).

\section{Conclusions}
\label{conclusions} 
In the paper, we considered six impostor partition functions $Z_i$, for $i=1-6$, obtained from the ``fake'' enumerator polynomials \eqref{eqn:fake} -- the polynomials satisfying all the properties of refined enumerator polynomials of real self-dual quantum stabilizer codes, yet not being  enumerator polynomials of any code. Then, $Z_i$ are obtained  using the explicit relation between  enumerator  polynomial  and code CFT torus partition function \eqref{eqn:recipe}, which in our case serves as an ansatz. We investigated these six $Z_i$ and denounced four of them as fake -- they are not torus partition functions of any unitary 2d CFT. Our analysis is inconclusive for the remaining two. We also ruled out the possibility of any of the six being Narain theories. 

We came to these conclusions by analysing the chiral algebra of the underlying CFTs, assuming that the $Z_i$ were genuine. The smallness of the central charge $c=3$, together with the presence of conserved currents imposed strict limitations on the possible chiral algebras, and for each case, we could reduce the choice to a handful of scenarios. Then, for each such scenario, we noticed that the number of  states of a particular dimension does not conform to the size of the representations of the would-be chiral algebra, leading to a contradiction. Our analysis is based on the specifics of each case. 

Ruling out $Z_i$ as the partition functions of Narain theories can be done separately, using the linearity of OPE coefficients. In this case, all the primary states of $\mathfrak{u}(1)^n \times \mathfrak{u}(1)^n$ algebra form a Narain lattice, hence a sum  of two states (in the sense of the $\mathfrak{u}(1)^n \times \mathfrak{u}(1)^n$ charges) must be a primary state as well.  Starting from the zero-spin states of small dimension, we can construct a zero-spin state of higher dimension, only to notice that such a state is not present in the spectrum of the would-be CFT. The analysis of the Narain case can be made more systematic. One can find all Euclidean lattices which reproduce the ``Euclidean'' spectrum of $Z_i$ with the purely imaginary $\tau$. This list is short: there are no such lattices for $Z_1,Z_2$, immediately ruling them out, and there is a unique lattice for each  $Z_i$ with $i=\overline{3,6}$. In the latter case, an analysis of the would-be Lorentzian structure is necessary to show that these Euclidean lattices equipped with the Lorentzian inner product of the most general form cannot reproduce $Z_i(\tau,\bar \tau)$ for complex $\tau$. 

Our analysis for $Z_{1,2}$ is incomplete; we could not rule them out as CFT partition functions or find the appropriate theories matching the spectrum. It would be interesting to complete this task as a first step to answer a more general question -- if the $Z$ obtained from a fake polynomial can ever be genuine, i.e.~be a partition function of an actual CFT. CFTs are vastly more rich and complicated objects than codes, and hence there are many more self-consistency conditions for $Z$.  Provided fake $W$ always lead to fake $Z$, there is therefore hope that these conditions can be used to analyse enumerator polynomials and identify the fake ones. The latter task is an open problem in coding theory, and is necessary to answer the following question -- if an optimal code of a particular length is also extremal. An attempt to make a step in this direction and to use the CFT consistency conditions coming from the higher genus partition functions was recently undertaken in  \cite{henriksson2021codes,henriksson2022narain}.

Our results are a confirmation that not all solutions of the torus modular bootstrap constraints are genuine CFTs. These are not the very first examples of this kind, there are known constructions built of chiral models \cite{schellekens1992meromorphic}, essentially providing examples of fake $Z$'s with one character. There is also an infinite class of candidate partition functions for rational CFTs with two characters \cite{mukhi2019classification}, many of which are expected to be fake. Our examples involve a handful of characters and also belong to an infinite discrete family, yielding  millions of possibly fake $Z$'s already for small central charge $c\leq 10$ \cite{dymarsky2021solutions}. In fact the construction based on fake polynomials can be generalized to produce {\it continuous} families of impostor $Z$; we construct simplest examples with $c=3$ in the appendix \ref{contfamily}.
Thus, our results emphasize the point that fake $Z$'s which solve modular bootstrap constraints -- yet not being partition functions of any CFT -- can be complicated, involve many characters and require an extremely elaborate analysis to rule out as fake. This picture is important to keep in mind because it contradicts the success of modular bootstrap to study particular theories of interest. Perhaps to understand why no fake $Z$ was found numerically, by solving a particular set of bootstrap constraints, one should address the following question --  can a $Z$, understood as a solution of modular bootstrap, which is sitting at the  boundary of an exclusion plot ever be fake? Our experience so far suggests that the answer is negative, but the underlying reason is unclear. 

 \section{Acknowledgements}
We thank Ofer Aharony for collaboration at the early stages of this project, and for comments on the draft. We also thank Felix Jonas and the Naama Barkai lab for help with computing, and Hiromi Ebisu, Masataka Watanabe, Ohad Mamroud, Adam Schwimmer, Shaul Zemel, Micha Berkooz, Erez Urbach, Adar Sharon, and Shai Chester for useful discussions.
AD is grateful to Weizmann Institute of Science for hospitality and acknowledges  sabbatical support of the Schwartz/Reisman Institute 
for Theoretical Physics.  AD was supported by the National Science Foundation under Grant No.~PHY-2013812. The work of RRK was supported in part by an Israel Science Foundation (ISF) center for excellence grant (grant number 2289/18), by ISF grant no. 2159/22, by Simons Foundation grant 994296 (Simons Collaboration on Confinement and QCD Strings), by grant no. 2018068 from the United States-Israel Binational Science Foundation (BSF), by the Minerva foundation with funding from the Federal German Ministry for Education and Research, by the German Research Foundation through a German-Israeli Project Cooperation (DIP) grant ``Holography and the Swampland'', and by a research grant from Martin Eisenstein. 
 
 \appendix
\section{Elementary proof for other \texorpdfstring{$Z_i$}{Zi} }
\label{elementaryproof}

\subsection{\texorpdfstring{$Z_1$}{Z1} }

\begin{align}
\tilde{Z}_1  ((\varrho t)^8 , (\varrho t ^{-1} )^8 )
	& = 
		1 + 4 \varrho ^2  + 12 \varrho ^4 + 8 \varrho ^6  + \dots + 24 \lb  t^{8}+t^{-8} \rb \varrho ^{14} + \dots   
\end{align}
We can proceed exactly as in the $\tilde{Z}_3$ case. We need at least two basis elements with sq-length 2 : $ a_{1,2} $; these satisfy $ a_1 . a_2 = 0 $ (see \eqref{eqn:polaa}). This means that $ \pm (a_1 \pm a_2 ) $ are the only linear combinations of $a_{1,2}$ with sq-length 4, and the remaining 8 are linearly independent of $a_{1,2}$. We select one of these as a basis element and call it $b_1$. \eqref{eqn:polab} tells us that $ a_{1,2} . b \in \{ 0, \pm 1 \} $, and we find that either choice leads us to a contradiction (see Fig. \ref{fig:Z13})

\subsection{\texorpdfstring{$Z_2$}{Z2} }

\begin{align}
\tilde{Z}_2 ((\varrho t)^8 , (\varrho t ^{-1} )^8 )
	& = 
		1 + 2 \varrho ^2  + 12 \varrho ^4 + 8 \varrho ^6 + 12 \varrho ^8 + \dots + 24 \lb  t^8 +t^{-8} \rb \varrho ^{14} + \dots
\label{eqn:Z2}
\end{align}
We select one of the sq-length 2 vectors : $a_1$, as a basis element, All sq-length 4 vectors with the exception of $\pm 2 a_1 $ are independent of $a_1$ -- we can therefore pick a new basis element $ b_1 $ of sq-length 4; \eqref{eqn:polab} gives us $ a_1 . b_1 \in \{  0, \pm 1 \} $ and we must choose $ a_1 . b_1 = 0 $ to avoid  \eqref{eqn:exContra}. Since no linear combination of $a_1 $ and $b_1 $ has sq-length 4, we must have another independent $ b_2 $ (with $ a_1 . b_2 = 0$). We run the polarisation identity again:
\begin{align}
|| b_1 + b_2 ||^2 + || b_1 - b_2 ||^2 & = 2 ( ||b_1||^2 + ||b_2||^2 ) \nonumber \\
                               & = 16   \nonumber \\
                               & = 4 + 12 = 6 + 10 = 8 + 8 = 10 + 6 = 12 + 4
                    \label{eqn:argsplit2}
\end{align}
to conclude that $ b_1 . b_2 \in \{ 0, \pm 1 , \pm 2 \} $. If $ b_1. b_2 = \pm 2 $,
\begin{align}
\boxed{
\begin{array}{rl}
    || a_1 + b_{1,2} ||^2   = 6 & \Rightarrow \langle a_1 , b_{1,2} \rangle = 0  \\
    || b_1 \mp b_2 ||^2  = 4 & \Rightarrow \langle b_1 , b_2 \rangle = 0 \\
    ||| a_1 + 2 b_1 \mp b_2 ||^2 = 14 & \Rightarrow 
                    \mp 2 \langle b_1 , b_2 \rangle + 2 \langle a_1 , b_1 \rangle \mp \langle a_1 , b_2 \rangle \neq 0
\end{array}
}
\label{eqn:bbm2}
\end{align}
If $ b_1. b_2 = \pm 1 $,
\begin{align}
\boxed{
\begin{array}{rl}
    || a_1 + b_{1,2} ||^2   = 6 & \Rightarrow \langle a_1 , b_{1,2} \rangle = 0 \\
    || b_1 \mp b_2 ||^2  = 6  & \Rightarrow \langle b_1 , b_2 \rangle = 0 \\
    ||| 2a_1 + b_1 \mp b_2 ||^2 = 14  & \Rightarrow 
                    \mp \langle b_1 , b_2 \rangle + 2 \langle a _1 , b_1 \rangle \mp 2 \langle a _1 , b_2 \rangle \neq 0
\end{array}
}
\label{eqn:bb1}
\end{align}
So $b_1.b_2 = 0$. Since $ (a_1, b_1, b_2 )$ are orthogonal, there are no non-trivial linear combinations with sq-length $ \leq 4 $ and we can find yet another independent $b_3$ ($a_1 . b_3 = 0 $, of course). We can recycle the arguments above to conclude that $ (a_1, b_1, b_2 , b_3 )$ are orthogonal. But then there must be at least 12 vectors with sq-length 6 ($ \pm a_1 \pm b_{1,2,3} $); this conflicts with the information presented in \eqref{eqn:Z2}.

\subsection{\texorpdfstring{$Z_5$}{Z5} }

\begin{align}
\tilde{Z}_5 ((\varrho t)^8 , (\varrho t ^{-1} )^8 )
	& = 
		1 + 2 \varrho ^2  + 4 \varrho ^4 + 24 \varrho ^6 + \lb 8 \lb  t^8 +t^{-8} \rb + 28 \rb \varrho ^8 + \dots + 72 \lb  t^8 +t^{-8} \rb \varrho ^{14} + \dots  
    \nonumber \\
  & \qquad + (96 ( t^{24} + t^{-24} ) + 216 ( t^{8} + t^{-8} ) ) \varrho ^{30} + \dots
\end{align}
Once again, the only vectors with sq-length 2 are $ \pm a _1 $. At sq-length 4, we have to bring in a new basis element $b_1$. We have to have $ a _1. b_1 =0 $ to avoid \eqref{eqn:exContra}; this tells us that the other sq-length 4 vectors are independent of $a_1, b_1$ -- let's call them $ \pm b_2 $. Again, $ a_1. b_2 = 0 $, and running the polarisation identity on $ b_{1,2}$ tells is that $ b_1. b_2 \in \{ 0, \pm 1 \}$.\footnote{This time, $b_1 . b_2 \neq \pm 2 $ because we are out of sq-length 4 vectors.}  If we pick the latter, we run into \eqref{eqn:bb1}. 

Thus, $(a_1, b_1, b_2) $ are orthogonal. We then have an independent $c _1 $ with sq-length 6, and:
\begin{align}
|| a_1 + c_1 ||^2 + || a_1 - c_1 ||^2 & = 2 ( ||a_1||^2 + ||c_1||^2 ) \nonumber \\
                               & = 16   \nonumber \\
                               & = 6 + 10 = 8 + 8 = 10 + 6
\end{align}
These correspond to $a_1 .c_1 \in \{ 0, \pm 1 \} $. Let's try out the second option: 
\begin{align}
\boxed{
\begin{array}{rl}
    || a_1 \mp c_1 || ^2 = 6 & \Rightarrow  \langle a_1 , c_1 \rangle = 0 \\
    || a_1 \pm 2 c_1 ||^2 = 30 & \Rightarrow  \langle a_1 , c_1 \rangle \neq 0 
\end{array}
}
\label{eqn:ac1}
\end{align}
Hence, we must have $a_1 .c_1 = 0 $. But then
\begin{align}
\boxed{
\begin{array}{rl}
    || 2 a_1 + c_1 ||^2 = 14 & \Rightarrow  2 \langle a_1 , c_1 \rangle = \pm 4 \\
    || a_1 + c_1 ||^2 = 8  & \Rightarrow  2 \langle a_1 , c_1 \rangle \in \{ 0, \pm 8 \} 
\end{array}
}
\label{eqn:ac0}
\end{align}

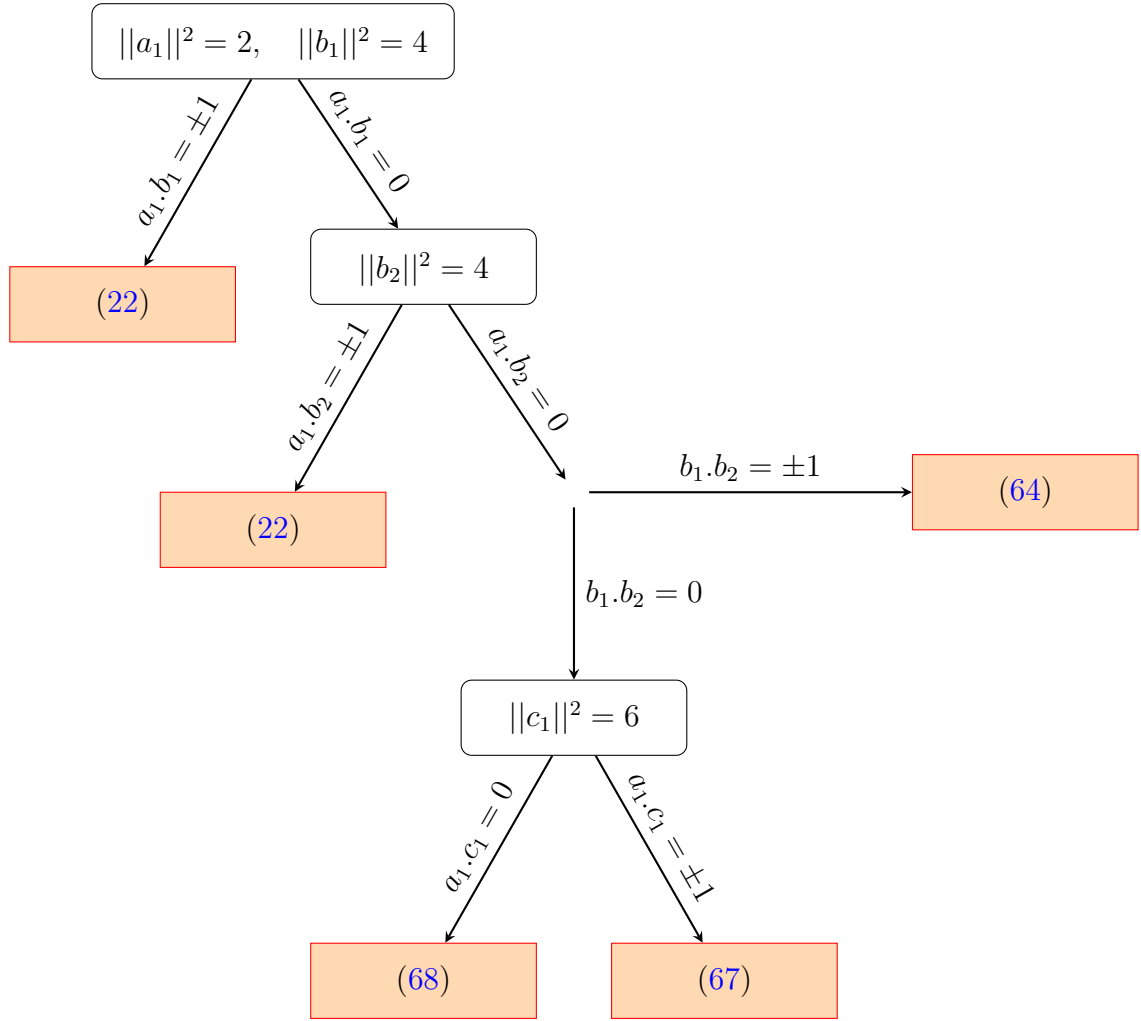
\begin{figure}

\begin{tikzpicture}[node distance=3cm]

    \node (start) [startstop] { \begin{tabular}{c}
                                     $ ||a_1||^2 = 2 , \quad ||b_1||^2 = 4 $  
                                \end{tabular}
                                };
    \node (ab1) [contra, below of=start, xshift=-2cm, yshift=-0.5cm] {\eqref{eqn:exContra}};
    
    \node (ab2) [startstop, below of=start, xshift=2cm] { \begin{tabular}{c}
                                                               $ ||b_2||^2 = 4 $  
                                                          \end{tabular}
                                                        };
    \node (ab3) [contra, below of=ab2, xshift=-2cm, yshift=-0.5cm] { \eqref{eqn:exContra} };
    \node (ab4) [circle, below of=ab2, xshift=2cm] {};
    \node (bb2) [startstop, below of=ab4, xshift=0cm] { $ || c_1 ||^2 = 6 $ };
    \node (bb3) [contra, below of=ab2, xshift=8cm] { \eqref{eqn:bb1} };

    \node (final1) [contra, below of=bb2, xshift=-2cm, yshift=-0.5cm] { \eqref{eqn:ac0} };
    \node (final2) [contra, below of=bb2, xshift=2cm, yshift=-0.5cm] { \eqref{eqn:ac1} };

    \draw [arrow] (start) -- node[above, sloped] {$ a _1. b_1 = \pm 1 $} (ab1);
    \draw [arrow] (start) -- node[above, sloped] {$ a _1. b_1 = 0 $} (ab2);
    \draw [arrow] (ab2) -- node[above, sloped] {$ a_1 . b_2 = \pm 1 $} (ab3);
    \draw [arrow] (ab2) -- node[above, sloped] {$ a _1. b_2 = 0 $} (ab4);

    \draw [arrow] (ab4) -- node[anchor=west] {$ b_1 . b_2 = 0 $} (bb2);
    \draw [arrow] (ab4) -- node[sloped, above] {$ b_1 . b_2 = \pm 1  $} (bb3);

    

    \draw [arrow] (bb2) -- node[sloped, above] {$ a_1.c_1 = 0$} (final1);
    \draw [arrow] (bb2) -- node[sloped, above] {$ a_1.c_1 = \pm 1 $} (final2);
    
\end{tikzpicture}

\caption{The outline of the argument in the case of $\tilde{Z}_5$. The coloured boxes indicate contradictions. }

\end{figure}

\subsection{\texorpdfstring{$Z_4$}{Z4}}

\begin{align}
\tilde{Z}_4 ( (\varrho  t )^8 , (\varrho t^{-1})^8 )
	& = 
		1 + 8 \varrho ^4  + 16 \varrho ^6 + \lb 4( t^8 +t^{-8} ) + 20 \rb \varrho ^{8}  +  \lb 16( t^8 +t^{-8} ) + 32 \rb \varrho ^{10} + \dots 
    \nonumber \\
&        \qquad
            + 48 \lb  t^8 +t^{-8} \rb \varrho ^{14} + \dots +   \lb  48 \lb t^{40} + t^{-40} \rb + 192 \lb t^8 + t^{-8} \rb + 288 \lb t^{24} + t^{-24} \rb  \rb \varrho ^{46}
    \nonumber \\
 &  \qquad     + \dots
            +   \lb  144 \lb t^{24} + t^{-24} \rb + 144 \lb t^{56} + t^{-56} \rb + 288 \lb t^{40} + t^{-40} \rb + 384 \lb t^8 + t^{-8} \rb  \rb \varrho ^{62} 
        \nonumber   \\
      & \qquad \qquad \qquad  + \dots 
\label{eqn:tilZ4} 
\end{align}

Our first step is to show that the sublattice generated by sq-length 4 vectors is $ \mathcal{M}_4 \simeq (2\mathbbm{Z})^4 $ ($ \simeq $ in the Euclidean sense). We start by running the polarisation identity on two linearly independent sq-length 4 vectors $b, b'$:
\begin{align}
|| b + b' ||^2 + || b + b' ||^2 & = 2 ( ||b ||^2 + ||b' ||^2 ) \nonumber \\
                               & = 16   \nonumber \\
                               & = 4 + 12 = 6 + 10 = 8 + 8 = 10 + 6 = 12 + 4
\label{eqn:polbb}
\end{align}
so $ b.b' \in \{ 0, \pm 1 , \pm 2 \} $. If $ b .b ' = \pm 1 $, we run into:
\begin{align}
\boxed{
\begin{array}{rl}
    || b \mp b' ||^2 = 6 &  \Rightarrow \langle b, b' \rangle = 0 \\
    || 3 b \mp b' ||^2 = 46 & \Rightarrow \langle b, b' \rangle \neq 0 
\end{array}
}
\label{eqn:bbm1}
\end{align}
and so $ b.b' \in \{ 0, \pm 2 \} $; this means that $ \frac{1}{ \sqrt{2} } \mathcal{M}_4 $ is an integral lattice generated by sq-length 2 vectors, and hence must be a direct sum of the root lattices -- $ A_m $, $D_m $, $E_m $, and $ \mathbbm{Z} ^m $. Since there are no $E_m$ lattices at dimensions below six, we can disregard them immediately; the $D_m $'s have too many vectors with sq-length 2 (aka the \textit{kissing number}); and the $ \mathbbm{Z} ^m $'s have norm 1 vectors. Therefore, we can restrict our attention to the $A_m$'s. Kissing numbers add under $\oplus $ and hence the only option with the correct kissing number and dimension is $ A _1 ^4 \simeq (\sqrt{2} \mathbbm{Z} )^4 $ -- thus, $ \mathcal{M}_4 \simeq (2 \mathbbm{Z} )^4 $.
\begin{table}\centering
\ra{1.5}
\begin{tabular}{@{}ccc@{}}\toprule
 Lattice   &  Kissing number & min. sq-length \\
\cmidrule{1-3}
$A_{m} $  & $m(m+1)$ & 2 \\
$D_{m\geq 3 }$   & $2m(m-1)$ & 2 \\
$\mathbbm{Z}^m$   & $2m$ & 1 \\
\bottomrule
\end{tabular}
\caption{A list of root lattices and their relevant properties. The $n$ that appears in the name of the lattice indicates the dimension. `min. sq-length' gives the sq-length of the shortest basis vector. The number of shortest vectors is the kissing number.}
\label{tab:rootLattices}
\end{table}

Next we show that any vector with sq-length 6 is orthogonal to $\mathcal{M}_4 $. Let's run the polarisation argument between two vectors $b$ and $c$ of sq-lengths 4 and 6 respectively\footnote{Note that $ || b \pm c ||^2 \neq 4 $ because $ b \pm c \in \mathcal{M}_4 \Rightarrow c \in \mathcal{M}_4  $, and this would mean that $  (2 \mathbbm{Z})^4 $ has a vector of sq-length 6. }:
\begin{align}
|| b + c ||^2 + || b - c ||^2 & = 2 ( ||b||^2 + ||c||^2 ) \nonumber \\
                               & = 20   \nonumber \\
                               & = 6 + 14 = 8 + 12 = 10 + 10 = 12 + 8 = 14 + 6
    \nonumber \\
                        b . c & \in \{ 0, \pm 1 , \pm 2 \}
        \label{eqn:polbc}
\end{align}
$b.c = \pm 2$ is ruled out via:
\begin{align}
        \boxed{
        \begin{array}{rl}
        || b \mp c ||^2 = 6 & \Rightarrow \langle b, c \rangle = 0 \\
        || 2 b \mp c ||^2 = 14 & \Rightarrow \langle b, c \rangle \neq 0 \\
        \end{array}
           }
           \label{eqn:bc2}
\end{align} 
$b.c = \pm 1$ may be ruled out too, but with a bit more labour. Firstly, this means that $ b \mp c $ is of sq-length 8, but cannot be one of the scalars since:
\begin{align}
        \boxed{
        \begin{array}{rl}
        \llangle b \mp c \rrangle ^2 = 0 & \Rightarrow \langle b, c \rangle = 0 \\
        || 4 b \mp c ||^2 = 62 & \Rightarrow \langle b, c \rangle \neq 0 
        \end{array}
           }
    \label{eqn:bc1}
\end{align} 
Thus, we must have $ 2 \langle b , c \rangle = \pm 8  $ (again, since all the spinning $\Delta =1 $ operators are spin-1). But then,
\begin{align}
        \boxed{
        \begin{array}{rl}
        || 4 b \mp c || ^2 & = 62 \\ 
         \llangle 4 b \mp c \rrangle ^2 & = \pm 32 \\
        \end{array}
           }
\end{align} 
which is a contradiction since there are no spin-4 operators with $\Delta = \frac{31}{4}$. Hence, $b . c = 0$, and we have our result. 

Since there are 16 vectors of sq-length 6, we must have two basis elements of sq-length 6 (we can't have more since we already have 4 basis elements with sq-length 4 and the lattice is six dimensional). This means that the sublattice generated by vectors of sq-length is two dimensional and has kissing number 16. The known bound in two dimensions is 6 (and is attained by the $A_2$ lattice; see \cite{SPLAG}), and hence this is a contradiction. Alternatively, $ || b + c ||^2 = 10 $ for any $ b ,c $ of sq-lengths 4 and 6 respectively, and this gives us a minimum of 128 vectors of sq-length 10 -- a quick look at \eqref{eqn:tilZ4} tells us that we only have 64.

\subsection{\texorpdfstring{$Z_6$}{Z6}}

\begin{align}
\tilde{Z}_6 ( (\varrho  t )^8 , (\varrho t^{-1})^8 )
	& = 
		1 + 4 \varrho ^4  + 24 \varrho ^6 + \lb 8( t^8 +t^{-8} ) + 28 \rb \varrho ^{8} + \lb 8( t^8 +t^{-8} ) + 16 \rb \varrho ^{10} + \dots + 
    \nonumber \\
&        \qquad
            72 \lb  t^8 +t^{-8} \rb \varrho ^{14}  + \dots  +   \lb 72 \lb t^{40} + t^{-40} \rb + 288 \lb t^8 + t^{-8} \rb + 432 \lb t^{24} + t^{-24} \rb  \rb \varrho ^{46} 
    \nonumber \\ 
 &  \qquad  
            + \dots +   \lb  216 \lb t^{24} + t^{-24} \rb + 216 \lb t^{56} + t^{-56} \rb + 432 \lb t^{40} + t^{-40} \rb  + 576 \lb t^8 + t^{-8} \rb   \rb \varrho ^{62}
            \nonumber \\
        & \qquad \qquad \qquad  + \dots
\label{eqn:Z6} 
\end{align}

It is easy to bootstrap the sublattice generated by vectors of sq-length 4 (say $ \mathcal{M}_4  $). There must at least be two independent vectors with sq-length 4 (let's call these $ b_1, b_2 $). The polarisation identity (see \eqref{eqn:polbb}) gives $ b_1 . b_2 \in \{ 0, \pm 1 \pm 2 \} $. Now, $ b_1 . b_2 = \pm 1 $ may be ruled out via \eqref{eqn:bbm1} and $ b_1 . b_2 = \pm 2 $ produces too many vectors of sq-length 4; hence, $ b_1 . b_2 = 0 $ and $ \mathcal{M}_4 \simeq (2 \mathbbm{Z} )^2 $.  

Now, we prove that any vector of sq-length 6 is orthogonal to $\mathcal{M}_4 $. For any $ b ,c $ of sq-length 4 and 6 respectively, \eqref{eqn:polbc} works the same way, and $ b . c \neq \pm 2 $ thanks to \eqref{eqn:bc2}. Say $ b . c = \pm 1 $; this means that $ b \mp c $ is of sq-length 8. We also need $ 2 \langle b, c \rangle = \pm 8 $ to avoid \eqref{eqn:bc1}. As in the $ \Z _4 $ case, we can now construct a spin-4 operator at $ \Delta = \frac{31}{4} $:
\begin{align}
        \boxed{
        \begin{array}{rl}
        || 4 b \mp c || ^2 & = 62 \\ 
         \llangle 4 b \mp c \rrangle ^2 & = \pm 32 \\
        \end{array}
           }
\end{align} 

This leads us to the conclusion that $b .c =0 $; subsequently $ || b + c ||^2 = 10$, and we must have at least $ 24 \times 4 = 96 $ vectors of sq-length 10; \eqref{eqn:Z6} immediately cries out in protest.

 \section{Constructing theta functions of sublattices}
\label{trick}
In this subsection, we discuss a trick that lets us compute the (Euclidean) theta functions of strings of $B$ matrices (as in \eqref{eqn:Bstring}) exactly. This will let us save a lot of time since computing these brute force can be very time-consuming. 

We start by noting that given a lattice generated by $M$, the lattice $ M B $ where $B$ is an integer-valued matrix with determinant 2 forms an index-2 sublattice of $M$. Furthermore, if $M$ is integral, we have the following result:
\begin{lemma}
Any index-2 sublattice $\mathcal{M}'$ of an integral lattice $ \mathcal{M} $ is of the form:
\begin{align}
\mathcal{M}' = \partial _v \mathcal{M} & := \left\{ l \in \mathcal{M} , \ v ( l) \in 2 \mathbbm{Z} \right\}
\end{align}
for some $ v \in \mathcal{M}^* / 2 \mathcal{M}^* $; where $ \mathcal{M}^* $ is the dual lattice of $\mathcal{M} $. Conversely, $ \partial _v \mathcal{M} $ is an index-2 sublattice of $ \mathcal{M} $ for any such $v \neq 0$.
\end{lemma}
In particular, $\mathcal{M}_k = \partial_{v_k} \mathcal{M}_{k-1}  = B_{i_1} ... B_{i_k}$ is an index-2 sublattice of $\mathcal{M}_{k-1} = B_{i_1} ... B_{i_{k-1}}$. 
The special tool that aids us in our quest is a generalised theta function:
\begin{align}
\Theta _{\mathcal{M} } \lb \xi _1 ^{v_1} \times \xi _2 ^{v_2} \times ... \times \xi _m ^{v_m}, \varrho \rb
	& = 
	\sum _{ r \in \mathcal{M} }  \xi _1 ^{r . v _1} 
	 \xi _2 ^{r . v _2}
	 ...
	 \xi _m ^{r . v _m}
	 \varrho ^{||r ||^2  /2 } 
 \end{align} 
The theta function of $\partial_v \mathcal{M} $ is given by:
 \begin{align}
\Theta _{ \partial _{v} \mathcal{M} } \lb \xi _1 ^{w_1} \times \xi _2 ^{w_2} \times ... \times \xi _{m-1} ^{w_{m-1}}, \varrho \rb
	& = 
	\frac{1}{2} \lb
		\Theta _{  \mathcal{M} } \lb \xi _1 ^{w_1} \times \xi _2 ^{w_2} \times ... \times \xi _{m-1} ^{w_{m-1}} \times 1 ^{v}, \varrho \rb
		\right. \nonumber \\
		& \qquad \qquad 
		+ \left. \Theta _{  \mathcal{M}} \lb \xi _1 ^{w_1} \times \xi _2 ^{w_2} \times ... \times \xi _{m-1} ^{w_{m-1}} \times (-1) ^{v}, \varrho \rb
		\rb
    \nonumber \\
	& = 
	\sum _{ \substack{ r \in \mathcal{M} \\ r.v \in 2 \mathbbm{Z}} }  \xi _1 ^{r . v _1} 
	 \xi _2 ^{r . v _2}
	 ...
	 \xi _{m-1} ^{r . v _{m-1}}
	 \varrho ^{||r ||^2  /2 } 
 \end{align} 
All we needed was the generalised theta function of $\mathcal{M}$ and the dual vector $v$. Now, $ \mathcal{M}_k $ itself is of the form $ \partial _{v_k } ... \partial _{v_1 } \mathbbm{Z}^n $. This means that all we're missing is the generalised theta function of $\mathbbm{Z}^n $:
\begin{align}
\Theta _{\mathbbm{Z}^n } \lb \xi _1 ^{w_1} \times \xi _2 ^{w_2} \times ... \times \xi _m ^{w_m}, \varrho \rb
	& = \prod _{i = 1} ^n 
	 \theta _3 \lb \lb \prod _{j = 1} ^m \xi _j ^{\lb w_j\rb_i} \rb , \varrho \rb 
	 \label{eqn:genthetaId}
 \end{align} 
Recall that $\theta _3 $ is a Jacobi theta function.

\section{Continuous family of impostor $Z$}
\label{contfamily}
The construction of \cite{ds,dymarsky2021solutions}, which maps quantum codes to particular Narain theories with the  fixed lattice $\Gamma$, see section \ref{NTintro}, was generalized in \cite{angelinos2022optimal} to include lattices $\Gamma$ of varying size. More precisely, starting e.g.~from a real self-dual stabilizer code of $F_2$, one constructs {\it full} enumerator polynomial $W(t,x,y,z)$ which gives rise to CFT partition function via
\begin{eqnarray}
Z=W(\psi_{00},\psi_{10},\psi_{11},\psi_{01}),\quad \psi_{ab}=\frac{1}{ |\eta|^2}\sum_{n,m} e^{i\pi \tau(r(n+a/2)+r^{-1}(m+b/2))^2-i\pi \bar\tau(r^{-1}(n+a/2)-r(m+b/2))^2},
\end{eqnarray}
where $r$ is an arbitrary parameter. When $r=1$, $\psi_{10}=\psi_{01}$ and we return to \eqref{eqn:recipe} upon a substitution $t\rightarrow x,\,  x,z\rightarrow z,\, y\rightarrow y$. As in the case of refined enumerator polynomials, there are {\it fake} full enumerator polynomials, which satisfy necessary algebraic identities -- insurance under $y\rightarrow -y$ and under 
\begin{equation}
    t\rightarrow \frac{t+x+y+z}{ 2},\,\, x\rightarrow \frac{t+x-y-z}{ 2},\,\,
    y\rightarrow \frac{t-x+y-z}{ 2},\,\, z\rightarrow \frac{t-x-y+z}{ 2},
\end{equation}
but not associated with any codes. There are no such polynomials for $c=1,2$ but already eighteen distinct fake polynomials for $c=3$,
\begin{eqnarray}
t^3 + t^2 x + t x^2 + x y^2 + 2 t x z + x^2 z + y^2 z, \\
 t^3 + 2 t^2 x + 2 t x^2 + x^3 + x y^2 + t z^2, \\
 t^3 + t^2 x + x y^2 + t^2 z + 2 t x z + x^2 z + t z^2, \\
 t^3 + t^2 x + t x^2 + x^3 + 2 x y^2 + 2 t z^2, \\
 t^3 + 2 t^2 x + 2 t x^2 + x^3 + t y^2 + x z^2, \\
 t^3 + t^2 x + t y^2 + t^2 z + 2 t x z + x^2 z + x z^2, \\
 t^3 + t^2 x + t x^2 + t^2 z + 2 t x z + y^2 z + x z^2, \\
 t^3 + t y^2 + x y^2 + 2 t x z + x^2 z + y^2 z + x z^2, \\
 t^3 + x y^2 + t^2 z + 2 t x z + y^2 z + t z^2 + x z^2, 
 \end{eqnarray}
 \begin{eqnarray}
 t^3 + x^3 + t y^2 + 2 x y^2 + 2 t z^2 + x z^2, \\
 t^3 + t^2 x + t x^2 + x^3 + 2 t y^2 + 2 x z^2, \\
 t^3 + x^3 + 2 t y^2 + x y^2 + t z^2 + 2 x z^2, \\
 t^3 + t x^2 + 2 t y^2 + 2 x^2 z + y^2 z + z^3, \\
 t^3 + 2 t x^2 + t y^2 + x^2 z + 2 y^2 z + z^3, \\
 t^3 + 2 t y^2 + t^2 z + 2 x^2 z + t z^2 + z^3, \\
 t^3 + 2 t x^2 + t^2 z + 2 y^2 z + t z^2 + z^3, \\
 t^3 + t y^2 + 2 t^2 z + x^2 z + 2 t z^2 + z^3, \\
 t^3 + t x^2 + 2 t^2 z + y^2 z + 2 t z^2 + z^3.
\end{eqnarray}
They give rise to eighteen continuous families of impostor $Z$, parameterized by $r$. For $r=1$ they reduce to six polynomials/fake $Z$ discussed in this paper.

\bibliographystyle{jhep}    
\bibliography{library}

\end{document}